\documentclass[12pt,letterpaper]{article}

\usepackage{setspace}

\usepackage{graphicx}

\usepackage{siunitx}

\usepackage[numbers,sort&compress,merge]{natbib}
\bibpunct[, ]{[}{]}{,}{n}{,}{,}
\makeatletter
\renewcommand\@biblabel[1]{(#1)}
\makeatother

\usepackage{textcomp}
\usepackage{amsfonts}
\usepackage{amsmath}
\usepackage{amssymb}
\usepackage{xcolor}
\newcommand{\n}{\nonumber \\}
\newcommand{\ee}{\text{e}}
\newcommand{\tns}[1]{\textbf{\textit{#1}}}
\newcommand{\tnsgrk}[1]{{\boldsymbol #1}}
\newcommand{\opr}[1]{{\hat {#1}}}
\newcommand{\bra}[1]{\langle{#1}|}
\newcommand{\ket}[1]{|{#1}\rangle}
\newcommand{\prj}[1]{\langle{#1}}
\newcommand{\ie}{\textit{i.e.},~}
\newcommand{\eg}{\textit{e.g.},~}
\newcommand{\etc}{\textit{etc.}}
\newcommand{\secref}[1]{section \ref{#1}}
\newcommand{\figref}[1]{Fig.~\ref{#1}}
\newcommand{\eqnref}[1]{eq.~(\ref{#1})}

\newcommand{\Figref}[1]{Fig.~\ref{#1}}
\newcommand{\Eqnref}[1]{Eq.~(\ref{#1})}

\newcommand{\eqnsref}[2]{eqs.~(\ref{#1}) and (\ref{#2})}

\newcommand{\tabref}[1]{Table \ref{#1}}
\newcommand{\Tabref}[1]{Table \ref{#1}}
\newcommand{\appref}[1]{appendix \ref{#1}}
\newcommand{\Appref}[1]{Appendix \ref{#1}}

\definecolor{mygreen}{RGB}{0,188,0}
\newcommand{\new}[1]{#1}
\newcommand{\reworked}[1]{#1}

\begin{document}

\title{Excitonic Coupled-cluster Theory}
\author{
  Yuhong Liu,
  Anthony D.~Dutoi$^{*}$ \\   
  {\footnotesize \textit{Department of Chemistry, University of the Pacific, Stockton, California 95211, USA}} \\
  $^*$\texttt{\small adutoi@pacific.edu}
}
\maketitle

\begin{abstract}
A variant of coupled-cluster theory is described here,
 wherein the degrees of freedom are fluctuations of fragments between internally correlated states.
The effects of intra-fragment correlation on the inter-fragment interaction are
 pre-computed and permanently folded into an effective Hamiltonian,
 thus avoiding redundant evaluations of local relaxations associated with coupled fluctuations.
A companion article shows that a low-scaling step may be used to cast
 the electronic Hamiltonians of real systems
 into the form required.
Two proof-of-principle demonstrations are presented here
\new{
 for non-covalent interactions.
}
One uses harmonic oscillators, for which accuracy and algorithm structure can be carefully controlled
 in comparisons.
The other uses small electronic systems (Be atoms)
 to demonstrate compelling accuracy and efficiency,
 also when inter-fragment electron exchange and charge transfer must be handled.
Since the cost of the global calculation does not depend directly on the correlation models
 used for the fragments, this should provide a way to incorporate
 difficult electronic structure problems into large systems.
This framework opens a promising path for building tunable, systematically improvable
 methods to capture properties of systems interacting with a large number of other systems.
The extension to excited states is also straightforward.
\end{abstract}

		\section{Introduction}

Starting a few decades ago, and continuing apace today, enormous progress is being made
 in performing useful chemical simulations by decomposing large quantum-mechanical systems
 into recoupled sub-systems.
The state of the art generally consists of embedding fragments into the electrostatic
 environments of their neighbors
 (with various approaches to the exchange interaction) 
\reworked{
 \cite{
 Kitaura:1999:FMO,
 Hirata:2005:ESEmbedding,
 Fedorov:2007:FMO,
 Kamiya:2008:EStaticEmbedding,
 Jacobson:2011:XPS,
 Huang:2011:PotFuncEmbedding,
 Manby:2012:ExactDFTEmbedding,
 Bulik:2014:DFTEmbedding,
 Dresselhaus:2015:WFembedDFT},
}
 or using a fragment-based decomposition of a reference wavefunction for subsequent electron-correlation calculations \cite{Jeziorski:1994:SAPT,Byrd:2015:MolClusterCCPT};
 these approaches may be taken in combination with schemes for configurational sampling and techniques for handling
 redundancy in periodic systems \cite{Nolan:2010:Hierarchical,Muller:2011:Incremental,Beran:2016:CrystalPolymorphRev}.
\new{
Similarly, many local correlation methods
 \cite{
  Forner:1985:LocalCorrelation,
  Stoll:1992:LocalCorrDiamond,
  Saebo:1993:LocalCorrelation,
  Schutz:2001:DomainCCSD,
  Maslen:2005:MP4TRIM,
  Subotnik:2006:SmoothLocalCCSD,
  Li:2010:ClusterInMolecule,
  Hattig:2012:PNOMollerPlesset,
  Kristensen:2012:DivideExpandConsolidate,
  Liakos:2015:PNOCClimits}
 have been developed that use orbitals that are localized (not necessarily to a fragment)
 to separate strong and weak correlations, generally neglecting or treating perturbatively the long-range electron correlation.
}
\new{
For modeling of generic physical phenomena in lattices,
 dynamical mean field theory \cite{Kotliar:2006:DynamicalMeanField} has been used with model Hamiltonians to represent the entanglement
 of different site states.
}
The literature chronicling the evolution of fragment-based
\new{
(and related)
}
 methods is vast, and has been reviewed
 several times 
\reworked{
 \cite{Gordon:2011:FMOReview,Richard:2012:FragUnifiedView,Collins:2015:EnergyBasedFragMethods,Raghavachari:2015:FragmentReview,Huang:2008:ElecStructSolids,Wesolowski:2015:EmbeddingReview}.
}

Notwithstanding all of this progress, a more favorable ratio of accuracy to computational cost is always desirable, in order to broaden
 the coverage of reliable simulations, especially if the phenomena under investigation hinge on small energy differences
 or require levels of electronic detail not afforded by presently applicable wavefunctions or density functionals.
There exists a need for an \textit{ab initio} scheme of recoupling fragments, which
 has both a well-defined progression towards exactness and a flexible scheme of approximations.
Systematic improvability is 
\new{
 important because it is
}
 the only way to rigorously demonstrate the reliability of a model in the context of a \textit{specific} problem.
\new{
Within this context, we desire a scheme that is capable of handling the short-range correlation that is important for chemistry,
 the long-range correlations that holds large systems together, and also the coupling between them
 (to the extent that local correlation affects dynamic polarizabilities).
Yet we also desire to separate the treatments of these components of correlation.
}

To this end, this article describes and tests a generalized variant of the coupled-cluster (CC) model.
The CC wavefunction is quite generic with respect to the degrees of freedom to which it can be applied
 (see, for example, vibrational CC \cite{Christiansen:2004:VCCformalism, Seidler:2010:VCCReview, Faucheaux:2015:HighOrderVCC}).
The operative paradigm in this work is to use state-to-state fluctuations
 of entire fragments, rather than individual particle or inter-particle coordinates.
The method is called excitonically renormalized CC (XR-CC), since such fluctuations form 
 the conceptual site basis from which Frenkel--Davydov
 excitons \cite{Frenkel:1931:Excitons,Davydov:1971:MolecularExcitons,May:2011:ChgNrgXfer,Sisto:2014:AbInitoExciton,Morrison:2015:AbInitioExciton}
 are built.
The potential power of this approach is rooted in restricting each fragment to the
 space of its lowest-energy \textit{internally correlated} states,
 thus folding intra-fragment correlations into an effective Hamiltonian and
 compressing the description of the most relevant part of the Hilbert space.

To elucidate this concept mechanistically, consider that the usual description of
 dispersion forces requires at least connected double substitutions,
 already exhausting the excitation level of conventional 
 CC theory with single and double substitutions (CCSD).
Correlation corrections to fragment polarizabilities appear only with
 higher substitutions,
 such as with full or perturbative inclusion of connected triples [CCSDT or CCSD(T), respectively].
These account for electronic relaxations accompanying local charge fluctuations.
In contrast, if up to connected double substitutions of electrons
 were to enter \textit{fragment} wavefunctions within an
 overall XR-CC wavefunction with up to connected \textit{dimer} fluctuations (XR-CCSD),
 this would include implicit contributions from
 connected four-electron substitutions.
Furthermore, the effective fragment interactions in the renormalized Hamiltonian will
 decay much faster in space than the bare Coulomb potential of the \textit{ab initio} Hamiltonian.
\reworked{
 For these reasons, it is plausible that accuracy better than conventional CCSD, for example,
 could be reached for a computational cost much below that of CCSD,
}
\new{
 particularly for non-covalently interacting fragments, such as we focus on in this article.
}

In a companion article, the electronic Hamiltonian for general chemical systems was decomposed by fragment,
 in precisely the manner necessary to apply the XR-CC model.
Charge-transfer and inter-fragment electron exchange were handled rigorously also when fragment orbitals overlap.
The decomposition is formally exact,
\new{
 (even permitting covalently bound fragments,
 though we are not convinced of the practicability of this),
}
 which allows for systematically improvable approximations.
Therefore, systematic studies of sources of error can be made, thus aiding in the development of judicious approximations.
Though the use of high-quality states for the fragments will impact the computational cost of constructing an effective Hamiltonian
 (scaling only quadratically with system size, however), the cost of a given XR-CC variant built upon it
 will depend only on the numbers of fragments 
\reworked{
 and the number of states admitted per fragment.
}
The level of approximation may also be tuned by the choice of electronic structure method for the fragments
 (possibly different for each fragment).

Fragment states and fragment-decomposed Hamiltonians have a long and continuing history in quantum chemistry.
Symmetry-adapted perturbation theory (SAPT) \cite{Jeziorski:1994:SAPT} is rooted in the early work of 
 of London, Axilrod, and Teller, among others, which begins by expressing intermolecular interactions
 as sums over fragment eigenstates \cite{Stone:2013:IntermolForces}.
The SAPT(CC) approach, in particular, handles intra-fragment correlation by applying a CC transformation to the fragment states \cite{Korona:2009:SAPTCC}.
Enforcing global antisymmetry for general systems in SAPT is unwieldy, however, and only rare
 applications beyond dimers are found \cite{Podeszwa:2007:SAPTDFT3Body}.
The recently proposed molecular cluster perturbation theory \cite{Byrd:2015:MolClusterCCPT}
 also divides the super-system Hamiltonian into fragment terms and interactions, but global antisymmetry is neglected.

In low-order perturbative approaches it is anyway unnecessary to transform the many-electron bases of the fragments
 (this can be incorporated into the non-iterative action of the Hamiltonian).
The motivations for proceeding beyond low-order perturbation theory are manifold, however.
To begin with, long-range induction and other cooperative effects can be substantial and occur
 at relatively high orders (\eg{4th}) \cite{Cui:2006:InductionFourthOrder,Lao:2016:LargeSystemMBE}.
In addition to this, one expects further errors if polarizabilities from mean-field descriptions of the fragments are used,
 since excitation energies control the ``stiffness'' of a charge distribution (perturbative denominators),
 and these are known to be quite sensitive to correlation level \cite{Riley:2010:IntermolForceDissect}.
There is also dynamical screening or cooperation between fluctuations,
 collectively known as many-body dispersion (a ``body'' is a fragment),
 which are missing at low orders of perturbation theory, and this has been shown to be important \cite{Podeszwa:2007:SAPTDFT3Body,Ambrosetti:2016:WavelikeVdW}.

Choosing the many-electron basis for each fragment will play a decisive role in the efficiency
 of an actual XR-CC calculation.
It is important then to mention the success of
 the density matrix renormalization group (DMRG)
 in iteratively
\reworked{
 optimizing renormalized
}
 sub-system bases for tensor-network states \cite{White:1992:DMRG, Chan:2011:DMRGReview, Hedegard:2015:DMRGdispersion}.
While this aspect is similar in spirit to XR-CC, the global wavefunction is very different in structure.
We might anticipate lesser success with XR-CC when near neighbors are highly entangled;
 however, a more efficient treatment of system-wide dynamical correlation (a known weakness of DMRG) should result
 from the exponential Ansatz, which separates connected (correlated) and disconnected (coincidental)
 simultaneous fluctuations.
The XR-CC approach also does not exclude iterative optimization of the fragment bases.
\new{
Fragment-based Hamiltonian renormalization in advance of DMRG has also recently been done under the name of active-space decomposition \cite{Parker:2014:ActiveSpaceDecompDMRG,Kim:2015:ActiveSpaceCovalent};
 however no field-operator-like resolution was developed, which we require to apply coupled-cluster CC theory to it (and we will also allow for charge transfer).
}
\reworked{
Finally, XR-CC could be thought of as an application of block-correlated CC theory \cite{Li:2004:BCCCtheory,Fang:2008:BCCCbondbreaking}
 to Hamiltonians that partition correlation by fragment (also having overlapping orbitals), rather than for isolating a few strong correlations within orthogonal subspaces of the molecule.
}

In this article, we provide the essential equations describing the XR-CC model.
Two methodological tests are then undertaken.
In one case, the super-system is composed of model ``molecules'' 
 that are internally constructed of coupled harmonic oscillators.
The ease of implementation and availability of the exact solution for that problem allow for rigorous comparisons of errors
 and computational cost for similarly structured conventional and excitonic algorithms.
To provide initial evidence that the promising results apply also to electronic systems,
 accounting for inter-fragment electron exchange and charge transfer,
 XR-CC is applied to chains of up to 100 Be atoms.

		\section{Theory}

Following the notation established in the companion article,
 upper-case latin letters are used for ascending-ordered tuples of integers,
 written as $I = (i_1, i_2, \cdots)$, and
 a subscript on an index $i_m$
 also restricts it to the states of fragment $m$.
Sets are denoted as $\{y_i\}$, containing all $y_i$ for each value of $i$ allowed by $y$;
 summations implicitly run over all allowed index values, as well.

	\subsection{Fluctuation Operators}

Consider a generic super-system composed of $N$ fragments.
Given a complete basis $\{\ket{\psi_{i_m}}\}$ for the many-body state space of each fragment $m$,
 we first assume that the super-system state space is completely spanned by
 a set of states $\{\ket{\Psi_I}\}$ of the form
 \begin{eqnarray}
  \ket{\Psi_I} = \ket{\psi_{i_1}\cdots\psi_{i_N}}
 \end{eqnarray}
 with $I = (i_1, \cdots i_N)$.
This notation is intended to imply, foremostly,
 that $\ket{\Psi_I}$ is tensor-product-like in structure.
By collecting the fragment-state labels into a single ket, 
 it is furthermore implied that
 that the state has proper inter-particle exchange symmetry (if any) among the primitive coordinates
 (\eg{electrons}).

We next assert the existence of a set of fluctuation operators $\{\opr{\tau}^j_i\}$,
 where the lower and upper indices of each operator identify two states $\ket{\psi_i}$ and $\ket{\psi_j}$ (possibly the same), which must belong to the same fragment.
The action of $\opr{\tau}_{i_m}^{j_m}$ onto basis state $\ket{\Psi_K}$ is defined as changing the state of fragment $m$ according to
 \begin{eqnarray}\label{action_definition}	
  \opr{\tau}_{i_m}^{j_m}\ket{\psi_{k_1}\cdots\psi_{k_m}\cdots\psi_{k_N}} 
                                        = \delta_{j_m,k_m}\ket{\psi_{k_1}\cdots\psi_{i_m}\cdots\psi_{k_N}}
 \end{eqnarray}
This is constructed to be reminiscent of a number-conserving pair of field operators onto a single-determinant electronic state,
 such that the null state results if the upper (``destruction'') index corresponds to an ``empty'' fragment state.
(The preference for positioning indices is clarified in the companion article.)
These operators have the following commutation relation by definition (further discussion in the companion article)
 \begin{eqnarray}\label{the_commutator}
  \big[\opr{\tau}^j_i,\opr{\tau}^l_k\big]
   = ~ \delta_{jk}\,\opr{\tau}^l_i ~-~ \delta_{il}\,\opr{\tau}^j_k
 \end{eqnarray}
The fact that the commutator of two fragment fluctuation has maximum fragment rank
 of one importantly guarantees the necessary auto-truncation of the Baker--Campbell--Hausdorff (BCH)
 expansion for the similarity-transformed Hamiltonian \cite{Helgaker:2002:PurpleBook}, discussed shortly.

The assumptions that (1) a set of tensor products of fragment states builds a complete basis for the super-system space,
 and (2) the asserted fluctuation operators are well-defined in that space,
 are points that need to be proven for different classes of systems.
For the oscillator-model fragments
 (closed systems with distinguishable coordinates),
 these assumptions are trivially valid. 
In the companion article, they are shown to be valid also for fragment-decomposed electronic systems.

	\subsection{Coupled-cluster Ansatz}

According to the forgoing definition of the fluctuation operators,
 any basis state $\ket{\Psi_I}$ may be reached from any other basis state $\ket{\Psi_J}$ via a string of $N$ (or fewer) fluctuation operators.
Combined with the assumption of completeness of this basis, it is then
 straightforward to show that an arbitrary super-system state 
 has a unique resolution (to within a normalization factor) in terms of the full $N$th-order CC (FCC) Ansatz applied to a reference state $\ket{\Psi_O}$
 conforming to $\prj{\Psi_\text{FCC}}\ket{\Psi_O} = 1$, as
 \begin{eqnarray}\label{CC_Ansatz}
  \ket{\Psi_\text{FCC}} = \ee^{\opr{T}} \ket{\Psi_O}
 \end{eqnarray}
We have hereby identified the tuple $O = (o_1,\cdots o_N)$ as special,
 in that $\ket{\psi_{o_m}}$ is taken to be the reference state of fragment $m$.
Notably, to the extent that these are correlated fragment states, 
 $\ket{\Psi_O}$ may already include a large fraction of correlation.

The operator $\opr{T}$
 consists only of fluctuations away from the reference, denoted $\opr{\tau}^{o_m}_{u_m}$ with $u_m\neq o_m$, referred to specifically as \textit{excitations}
 \begin{eqnarray}\label{T_operator}
  \opr{T} = \sum_m\sum_{u_m\neq o_m} t_1^{u_m} ~ \opr{\tau}^{o_m}_{u_m}
          ~~ + \sum_{m_1<m_2}\sum_{\substack{u_{m_1}\neq o_{m_1}\\u_{m_2}\neq o_{m_2}}} t_2^{u_{m_1} u_{m_2}} ~ \opr{\tau}^{o_{m_1}}_{u_{m_1}} \, \opr{\tau}^{o_{m_2}}_{u_{m_2}}
          ~~ + \cdots
 \end{eqnarray}
In this case, single-excitation amplitudes $t_1^{u_m}$ are associated with monomers, and doubles amplitudes $t_2^{u_{m_1} u_{m_2}}$ are associated with dimers, {\etc}
The notation $m_1$$<$$m_2$ under a summation runs over all unique pairs ({\etc}) of fragments.
As with conventional CC theory, excitation operators all commute with one another.
We will henceforth reserve the indices $u_m$, $v_m$, $\cdots$
 to denote any many-electron state other than the reference, and leave the indices $i_m$, $j_m$, $\cdots$ as general.
These index letters are chosen to be reminiscent of ``occupied/zeroth-order'' and ``unoccupied/virtual.''

	\subsection{Hamiltonian}

With a general wavefunction Ansatz available, the central task is to
 iteratively determine the amplitudes $t_1^{i_m}$, $t_2^{i_{m_1}i_{m_2}}$, \etc,
 that approximate the ground state of a Hamiltonian $\opr{\mathcal{H}}$.
More precisely, the residual of the eigenstate condition must lie outside the space of variations considered.
This involves the familiar step of resolving the action of the similarity-transformed Hamiltonian
 $\ee^{-\opr{T}}\opr{\mathcal{H}}\ee^{\opr{T}}$
 onto the reference state.
In order to avoid expensive recourse to the primitive degrees of freedom during the iterations,
 $\opr{\mathcal{H}}$ must itself be written as an expansion in terms of strings of the fluctuation operators
 \begin{eqnarray}\label{H_operator}
  \opr{\mathcal{H}} = \sum_m\sum_{i_m,j_m} H^{i_m}_{j_m} ~ \opr{\tau}^{j_m}_{i_m}
                    ~~ + \sum_{m_1<m_2}\sum_{\substack{i_{m_1},j_{m_1} \\ i_{m_2},j_{m_2}}}
                         H^{i_{m_1} i_{m_2}}_{j_{m_1} j_{m_2}} ~ \opr{\tau}^{j_{m_1}}_{i_{m_1}} \, \opr{\tau}^{j_{m_2}}_{i_{m_2}}
                    ~~ + \cdots
 \end{eqnarray}
The elements $H^{i_m}_{j_m}$ build a Hamiltonian matrix for fragment $m$, and the higher-order terms are responsible for
 couplings between fragments (up to $N$th order, in principle, depending on the kind of system).

We have hereby added a third assertion, that the system Hamiltonian can be written in terms of fragment fluctuations.
(The original two were basis completeness and existence of fluctuation operators.)
It is likely possible to prove that
 these assertions are fulfilled
 for broad classes of systems, relying on only benevolent assumptions.
However, while interesting, it would be useless without an explicit form for
 the expansion of $\opr{\mathcal{H}}$ and computational recipes for the matrix elements therein.
For the oscillator-model fragments in this article, this will be trivial,
 but, for electronic systems that may overlap and transfer charge,
 possibly also having linear dependencies in the one-electron basis,
 the exercise is more intense, and it is undertaken in the companion article.

	\subsection{Correlation Models}

In approximate XR-CC methods, the expansion of $\opr{T}$ will be truncated at finite fragment order.
The established nomenclature for the CC hierarchy (CCSD, CCSDT, \etc) is hereby adopted
 to reference the included orders of connected fragment fluctuations.
An interesting analogue to a self-consistent-field (SCF) calculation, which would capture long-range induction using polarizabilities from
 correlated levels of theory, would be the use of only single excitations in $\opr{T}$ (XR-CCS).
Models beyond XR-CCS (\eg{XR-CCSD}) introduce entangled fluctuations among internally correlated fragment states,
 accounting for dispersion forces, {\etc},
 in a manner that is both self-consistent and size-consistent.

In addition to the correlation level of the wavefunction,
 there are approximations which may be applied to the excitonic Hamiltonian.
It is most sensible to include such qualifiers with the prepended ``XR.''
In this work, we will specifically refer to the methods XR2-CCSD and XR2-CCSDT,
 and also to the analogue of full configuration interaction (FCI), XR2-FCI.
The ``2'' in this declension indicates that the Hamiltonian contains maximally dimer coupling terms.
For electronic systems, this is an approximation because trimer and tetramer terms are missing;
 these are discussed in the companion article.

The central approximation in the XR-CC model is that the fragment state spaces are to be truncated according to the desired balance of
 cost against accuracy for the property under consideration.
Other parameters could then be given with the method designation, in order to indicate the numbers and qualities of fragment states used, \etc,
 but we withhold discussion of this until system details are presented.

	\subsection{Amplitude Equations \label{scaling}}

The final step is the theoretically straightforward task of using
 the BCH expansion and
 the commutator of \eqnref{the_commutator}
 to evaluate
 $\ket{\Omega} = \ee^{-\opr{T}}\opr{\mathcal{H}}\ee^{\opr{T}}\ket{\Psi_O}$ 
 in terms of excitations above the reference
 \begin{eqnarray}\label{omega_abstract}
  \ket{\Omega} = \Big[ \omega_0 ~+~ \sum_m \sum_{u_m} \omega_1^{u_m} ~ \opr{\tau}^{o_m}_{u_m}
          ~~ + \sum_{m_1<m_2} \sum_{u_{m_1},u_{m_2}} \omega_2^{u_{m_1} u_{m_2}} ~ \opr{\tau}^{o_{m_1}}_{u_{m_1}} \, \opr{\tau}^{o_{m_2}}_{u_{m_2}}
          ~~ + \cdots \Big]\ket{\Psi_O}
 \end{eqnarray}
Within a suitably chosen non-linear optimization algorithm,
 this similarity-transformed state-space energy gradient
 defines the update to $\opr{T}$,
 and $\omega_0$ is the pseudo-energy at any given iteration.

%

Regarding the concrete derivation of working amplitude equations,
 there are aspects that are both more and less complex than the conventional CC equations.
On the one hand, the Hamiltonian may contain up to four-fragment (four-body) interactions, and Wick's theorem
 would need to be generalized for two-index fluctuation operators in a generalized diagrammatic approach.
On the other hand, both indices of any operator reference two states of the same fragment, and no operator string in the Hamiltonian
 or cluster operator contains more than one fluctuation operator for any given fragment,
 making the algebraic approach tenable.

For completeness, the amplitude equations implemented here are given in \appref{amplitudes}, using the notation of this article.
However, these could have been
 borrowed from previous work on vibrational CC (VCC) theory \cite{Christiansen:2004:VCCimplementation,Seidler:2008:VCCefficient,Seidler:2009:VCCeqnsAutomatic},
 which will be particularly advantageous for expediting efficient XR-CC implementations in future work.
According to \eqnref{final_equations}, the XR2-CCSD method formally scales with the third power of the system size
  and the fourth power of the number of states included per fragment,
  which is indeed the same as VCC[2] with two-mode couplings
  (VCC[$n$] includes up to $n$-fold connected substitutions) \cite{Seidler:2009:VCCeqnsAutomatic}.
From this we know that XR3-CCSD and XR4-CCSD will have the $N^3$ and $N^4$ formal scalings of three- and four-mode VCC[2], respectively.
The isomorphism between VCC and XR-CC occurs because both map onto generic CC for distinguishable coordinates.
In this mapping,
 the many-electron states of a fragment correspond to the one-body states of some fictitious particle or mode.
Pairs of field operators that annihilate and create such single particles in those abstracted states
 serve as mappings for the two-index fluctuation operators.
As a matter of academic interest, in the Supplementary Information, we derive the amplitude equations
 using only the commutator of \eqnref{the_commutator}, without reliance on such a mapping.

		\section{Proof-of-principle Tests}

The two kinds of systems explored were chosen to 
 to provide initial evidence that the XR-CC formalism holds promise,
 while offering peculiar ease of implementation and/or special analytical properties.
Handling two very different kinds of systems also highlights the generality
 of CC theory here.

In the first systems, dipole-coupled fragments are groups of internally coupled harmonic oscillators.
These systems provide both trivial primitive matrix elements and the opportunity to compare accuracy against an exact solution.
The particular ease of implementation of the Hamiltonian allows us to compare
 XR-CC accuracy and timings to an algorithm that operates on primitive oscillator coordinates,
 but with precisely the same structure and level of optimization.
This is intended to mimic a directly comparable conventional CC algorithm,
 in order to isolate the effect of excitonic renormalization of the Hamiltonian.

In the second set of systems, the fragments are beryllium atoms.
Although the one-electron basis set is too small to be chemically meaningful,
 we nonetheless demonstrate the applicability of XR-CC to electronic
 systems, which require accounting for exchange antisymmetry and inter-fragment charge transfer.
Failing an exact solution for more than just a few Be atoms,
 we compare our accuracy to conventional CCSD and CCSD(T),
 which are also referenced in comparative timings.

All calculations in this work were run on a single 2.6 GHz Intel Xeon (E5) core
 of a dedicated node that was kept free of other traffic,
 with all runtime data kept in 2133 MHz core memory.
A completely in-house implementation of XR-CC methodology is used for experimental calculations;
 evaluation of the amplitude equations is coded in the C programming language and called as a shared library from
 a generic Python-based quasi-Newton driver, accelerated by direct inversion in the iterative subspace (DIIS) \cite{Helgaker:2002:PurpleBook}.
\Appref{code} discusses some further details concerning two independent implementations
 of the XR2-CCSD amplitude equations (one abstract, one efficient) and validation of the computer codes against standard packages.

	\subsection{Model Oscillator ``Molecules''}

\subsubsection{Systems and methods}

This study considers an \textit{a priori} pairwise fragment Hamiltonian of the following form,
 which is for generic closed systems of distinguishable internal coordinates,
 distributed in one dimension, and coupled only by the longitudinal dipole--dipole interaction
 \begin{eqnarray}\label{oscillatorH}
  \opr{\mathcal{H}} = \sum_m \opr{H}^{(m)} + \sum_{m_1<m_2} k^{(m_1,m_2)} \opr{\mu}^{(m_1)}\opr{\mu}^{(m_2)}
 \end{eqnarray}
$\opr{H}^{(m)}$ is the Hamiltonian of fragment $m$ in isolation, and $\opr{\mu}^{(m)}$ is
 its dipole operator along the super-system axis.
The coupling constant
 $k^{(m_1,m_2)} = -(2 E_\text{h} a_0)/(e^2 R_{m_1 m_2}^3)$
 depends on the distance $R_{m_1 m_2}$ between fragments $m_1$ and $m_2$.
$E_\text{h}$, $a_0$, and $e$ are atomic units of energy, length and charge, respectively.
Since the degrees of freedom are distinguishable, the tensor product set $\{\ket{\Psi_I}\}$,
 built from orthonormal sets of fragment states $\{\ket{\psi_{i_m}}\}$, is automatically an orthonormal basis for the super-system space.
Considering separately the cases where, one, two, \etc, fragment states differ
 between the bra and ket of a matrix element,
 it is straightforward to show that the Hamiltonian projected into this super-system basis may
 be rewritten exactly as
 \begin{eqnarray}\label{projoscH}
  &~& \opr{\mathcal{H}} = \sum_m\sum_{i_m,j_m} H^{i_m}_{j_m} ~ \opr{\tau}^{j_m}_{i_m}
                    ~~ + \sum_{m_1<m_2}\sum_{\substack{i_{m_1},j_{m_1} \\ i_{m_2},j_{m_2}}}
                         k^{(m_1,m_2)} \mu^{i_{m_1}}_{j_{m_1}} \mu^{i_{m_2}}_{j_{m_2}} ~ \opr{\tau}^{j_{m_1}}_{i_{m_1}} \, \opr{\tau}^{j_{m_2}}_{i_{m_2}} \n
  &~& \;\;\;\;\;
  H^{i_m}_{j_m} = \bra{\psi_{i_m}} \opr{H}^{(m)} \ket{\psi_{j_m}}
  \;\;\;\;\;
  \;\;\;\;\;
  \mu^{i_m}_{j_m} = \bra{\psi_{i_m}} \opr{\mu}^{(m)} \ket{\psi_{j_m}}
 \end{eqnarray}
 which intuitively has fragment rank of two.
This Hamiltonian is in the form that we have demanded for the XR-CC scheme.

It will be convenient to let each fragment be a group of linearly coupled harmonic oscillators,
 as a proxy for general internal interactions of fragment coordinates.
If we further consider the coordinate of each primitive oscillator as representing the distance 
 between two opposing charges along the direction of the inter-fragment axis,
 then the dipole operators of the fragments are  consequently defined.
Such a model is convenient because
 the overall system is then completely described by linear couplings between all pairs of primitive harmonic oscillators.
Diagonalization of the system-wide matrix of coupling constants can be used to efficiently obtain
 the ground-state energy to within machine precision, for purposes of comparison.
It should be made clear, however, that this method of exact solution
 relies on a special structure to the problem that is lost upon projection into a basis.
The dimension of the basis-projected super-system Hamiltonian scales exponentially with $N$,
 and, furthermore, sparsity due to the dipolar selection rules is lost  upon transformation to an internally correlated excitonic basis.
Such systems are then reasonable mimics for the complexity of systems of real molecules.
A Hamiltonian for weakly interacting molecules may indeed be written in the form of \eqnref{projoscH}.

Our test systems of this type consist of linear chains of 2 to 30 oscillator-model ``molecules,'' equally spaced by either 5 or 10 $a_0$.
Each such fragment consists of 8 internal oscillators
 with force constants spread evenly 
 over the range of 1 to 2 $E_\text{h}/a_0^2$, inclusive.
Each harmonic potential contains a particle with the same mass as an electron, and its coordinate is interpreted as
 the displacement of a $-e$ charge, relative to one of $+e$.
The coupling constant between each pair of internal oscillators has a positive (repulsive) value
 whose magnitude is 1/3 of the difference between their force constants, such that oscillators with even quite
 different force constants are substantially mixed in the fragment ground states.
For the XR-CC calculations we choose the set of internally correlated basis states $\{\ket{\psi_{i_m}}\}$
 for each fragment to be the set of exact energy eigenstates for each fragment in isolation,
 using the ground state of each fragment as its reference in the XR-CC wavefunction.
These fragment states, and the exact values of the corresponding matrix elements of $\hat{H}^{(m)}$ and $\hat{\mu}^{(m)}$,
 are available via diagonalization of the \textit{internal} coupling matrices for the individual fragments.

The accuracy and computational effort of the XR-CC calculations will be compared 
 to a CC calculation that operates on fluctuations of the primitive oscillators underlying the fragments.
This is intended to mimic the conventional practice of working with creation and annihilation operators for
 individual electrons.
Such a calculation is  performed here as an XR-CC calculation,
 but where each primitive oscillator is mapped to a fragment.
An important aspect of this
 is that the same computer program can be used for both the excitonic and conventional CC calculations.
Any efficiency advantage of XR-CC
 can then be traced directly to the renormalization of the Hamiltonian.

Given that only relative timings are important, the slower abstract implementation
 of the XR2-CCSD amplitude equations suffices for this study
 (see \appref{code}).
Since the Hamiltonian is manifestly pairwise, we suppress the ``2'' for the remainder of
 the discussion of these systems, as it does not constitute an approximation.
In the limit that each primitive oscillator is nominally a fragment, the ``X'' is also suppressed,
 such that CCSD refers here to the conventional variant.
For all calculations, convergence is declared when the energy correction changes by less than 1 part in $10^8$ between amplitude iterations.

\subsubsection{Results}

\begin{figure}
  \centering
  \includegraphics[width=7.5cm]{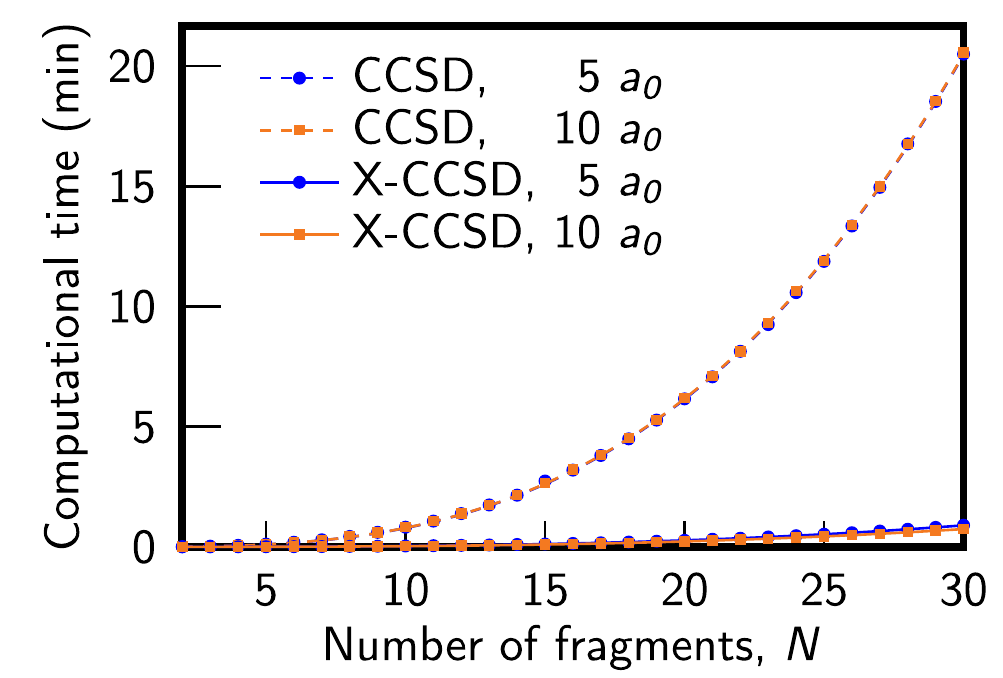}
  \caption{\label{oscillators}
   For either spacing between the oscillator-model``molecules,'' the computational time for XR-CCSD is much shorter than for
    an identically structured conventional CCSD algorithm.
   The best-fit monomial connecting curves are described well as cubic functions.
   \new{
   (The timings for either variant are nearly independent of the separation used, placing them almost on top of each other.)
   }
  }
\end{figure}

\begin{table}
  \begin{center}
    \begin{tabular}{c|c|c|c|c}
      \hline
      ~            & \multicolumn{2}{c|}{5 $a_0$}            & \multicolumn{2}{c}{10 $a_0$}           \\
      \hline
      ~            & best-fit timing & asymptotic error & best-fit timing & asymptotic error \\
      \hline
      CCSD    & $0.051 \times N^{2.97}$ & $8.3 \times 10^{-4}$   &   $0.052\times N^{2.96}$ & $8.2 \times 10^{-4}$  \\
      XR-CCSD & $0.0025\times N^{2.94}$ & $1.4 \times 10^{-6}$   &   $0.0021\times N^{2.93}$ & $3.2 \times 10^{-10}$ \\
      \hline
    \end{tabular}
  \end{center}
  \caption{\label{summary}
    XR-CCSD is both more accurate and faster than conventional CCSD, for identically structured algorithms.
    For each spacing between the oscillator-model ``molecules,'' the best-fit monomial for the timing curve (in sec) is given along with the 
     absolute error for large $N$ (in $E_\text{h}$/fragment).
    Since XR-CCSD does not need to account for internal fragment correlations, the error decays to zero with separation.
  }
\end{table}

The computation times for CCSD and XR-CCSD are plotted against the number of fragments in \figref{oscillators}, for each inter-fragment spacing.
All reported timings are for the smallest (fragment many-body) basis that converges to the inherent accuracy limit
 of the finite-order (XR-)CCSD model (\ie{the complete-basis limit});
 the determination of this convergence is discussed in \secref{osc_details}.

The excitonic calculations are seen to be much faster than the conventional variant, as reflected in the 
 best-fit monomials for each series, recorded in \tabref{summary} (log--log least-squares regression for the largest 15 systems),
 which are consistent with our expectation of third-order scaling
\new{
 (see \secref{scaling}).
}
The 20-fold decrease in the prefactor for XR-CCSD can be rationalized from the previously noted scalings
 as $(1/8)^3(9/4)^4$ to account for the fact that the same algorithm is operating on 1/8 as many ``fragments''
 (a ``fragment'' is a primitive oscillator for CCSD), but requiring 9 states per fragment, instead of 4 states per oscillator, as discussed in \secref{osc_details}.

\Tabref{summary} also gives the absolute error for the respective basis-converged calculations of the 30-fragment systems.
This shows the excitonic variant to be not only faster, but also better.
It is 3 orders of magnitude more accurate at 5 $a_0$ separation,
 and the error decays to zero with separation, since XR-CC only computes the interaction energy.

\subsubsection{Procedural details \label{osc_details}}

Determination of convergence with respect to basis size was assessed by observing the approach to the inherent method error
 as more states of increasing energy were considered for each sub-system (each fragment or each primitive oscillator).
This is facilitated by our knowledge of the exact ground-state energy for these special systems.
The limiting method error is due to the truncated excitation level in the (XR-)CCSD wavefunction;
 it is different for the excitonic and conventional variants, and it depends on the inter-fragment spacing.
Expressed as a fraction of the energy difference $\Delta E$ that
 separates the exact ground state from the relevant (excitonic or conventional) reference,
 the method error is fairly independent of system size for the larger $N$.
An asymptotic estimate of this constant unrecovered fraction of $\Delta E$ was obtained by monitoring the error as a function of basis size
 for the 30-fragment system
 and declaring convergence when the order of magnitude and two significant figures of its ratio with $\Delta E$ were stable.
A few percent increase over this value was then established as a convergence threshold for this fraction,
 used for convenience to signal basis-set convergence for all other system sizes (of same spacing and CC variant).
Irrespective of spacing or number of fragments, application of the aforementioned criterion uniformly demanded
 9 states per fragment for the XR-CCSD calculations  and
 4 states per oscillator for the CCSD calculations (\ie{32 states per fragment}).

\begin{table}
  \begin{center}
    \begin{tabular}{c|c|c|c|c}
      \hline
      ~            & \multicolumn{2}{c|}{5 $a_0$}            & \multicolumn{2}{c}{10 $a_0$}           \\
      \hline
      ~            & exact $\Delta E$ & thresh.~fraction & exact $\Delta E$ & thresh.~fraction \\
      \hline
      CCSD    & $-5.0\times 10^{-2}$ & $1.7 \times 10^{-2}$ & $-4.9\times 10^{-2}$ & $1.7 \times 10^{-2}$ \\
      XR-CCSD & $-5.5\times 10^{-4}$ & $2.6 \times 10^{-3}$ & $-8.5\times 10^{-6}$ & $4.2 \times 10^{-5}$ \\
      \hline
    \end{tabular}
  \end{center}
  \caption{\label{thresholds}
    For the oscillator-model ``molecules,'' the energy change $\Delta E$ (in $E_\text{h}$/fragment) to reach the ground state,
     from either a conventional (CCSD) or excitonic (XR-CCSD) reference state, is given for large $N$.
    The threshold fraction of $\Delta E$ that is unrecovered near convergence to the complete basis limit is also given.
    XR-CCSD recovers a larger fraction of an already smaller energetic distance to the ground state.
  }
\end{table}

\Tabref{thresholds} gives the per-fragment value of the effectively asymptotic $\Delta E$ (for 30 fragments)
 for each spacing and reference state.
This value is much larger in the conventional case, where it represents the entire correlation energy,
 as opposed to only the inter-fragment interaction energy;
 the excitonic value 
 also
 asymptotes to zero with increased separation.
Also given in \tabref{thresholds} are the aforementioned threshold error fractions.
In addition to the value of $\Delta E$ being smaller when starting from an excitonic reference,
 XR-CCSD leaves a smaller fraction of this already smaller energetic distance to the ground state unrecovered.
Note that the product of each pair of $\Delta E$ and threshold-fraction value in \tabref{thresholds} is, in fact, a slight overestimate of the
 respective absolute error reported in \tabref{summary}.

	\subsection{Beryllium-atom Chains}

\subsubsection{Systems and methods}

This study performs tests of the XR2-CCSD model using simple electronic fragments.
These are Be atoms in a 6-31G basis, with the core electrons frozen in the Hartree--Fock 1s orbitals of the isolated atoms.
These fragments are then effectively 2-electron problems (when neutral).
%
The basis set here is also too small for chemically meaningful results.
However, our purpose is only to set up a model problem that has all the essential
 features of fragment-decomposed electronic systems: inter-fragment electron exchange, charge transfer,
 and strong intra-fragment correlations.

It is worth noting for context that the Be dimer is already a notoriously difficult multi-reference problem,
 which has received much attention over several
 decades \cite{
 Ewing:1970:MetalDimers,
 Blomberg:1978:Be2,
 Bondybey:1984:TheoryExptBe2,
 Kowalski:2005:Be3,
 Merritt:2009:Be2experimental,
 El_Khatib:2014:Be2StaticCorrelation,
 Sharma:2014:Be2DMRG,
 Meshkov:2014:Be2potential}.
This is ultimately due to the fact that the two valence electrons of each 
 atom live in a space of four nearly degenerate orbitals;
 this leads to substantial angular correlation (evident even in a 6-31G basis \cite{Dutoi:2004:Radicals}).
The dimer bonding curve has a long-appreciated odd shape, which illuminatingly separates into two distinct minima
 as the basis size is decreased from cc-pVTZ to cc-pVDZ
 in FCI calculations.
The deeper inner minimum near 2.5 {\AA} eventually converges with basis quality to the observed dissociation depth of roughly 4 m$E_\text{h}$ (10 kJ/mol).
The shallower outer minimum near 4.5 {\AA} 
 merges in to form a noticeable shoulder of the binding well.
Only this outer minimum is captured using the 6-31G basis.

Unlike for the study of the oscillator ``molecules,'' 
 the basis for the primitive (one-electron) coordinates is a fixed feature of these simulations.
Accuracy here is then hypothetically defined relative to the FCI solution of the
\reworked{
 finite-basis
}
 problem.
Although the one-electron basis is removed as a variable from this discussion,
 one of the most important remaining considerations is the many-electron basis used for each atomic fragment
 in the XR-CC calculations.
Indeed, the decision to use such a small one-electron basis is 
 driven by the desire 
 to nimbly explore this crucial facet.

There are two easily anticipated difficulties with the conceptually simplest scheme of using fragment eigenstates as the many-electron basis,
 even assuming that FCI on the fragments were not an issue.
These were indeed both met in practice.
The first is that prioritizing states on the basis of their energy alone
 ignores the equally important consideration of the coupling strengths;
 some low-energy atomic states will be unimportant to inter-fragment interactions, whereas some high-energy states will couple strongly.
The second is that, even for nominally non-covalent interactions, some charge transfer will be important;
 consider the quintessential example of the semi-covalent hydrogen bond \cite{Khaliullin:2009:WaterDimerCT}.
Recovering this interaction would involve converging also cationic and anionic eigenstates of the fragments.
The anionic states will be particularly problematic if they are not stable.
However, in the field of a nearby electron hole, such states can contribute.
As an easily accessible example, consider that the Li$_2$ single bond
 is described by substantial two-center electron resonance,
 in spite of the small electron affinity of Li.

A reasonable manner in which to circumvent both of the forgoing issues, at least theoretically,
 is to work ``backwards,'' identifying those atomic states of greatest import for the
 nearest-neighbor interaction, and then using
 these states 
 for the global problem.
The essential tool for this here is the construction of a density matrix in the Fock space
 for one atom in a dimer.
Working with such small fragments, we can do this by brute force,
 decomposing the FCI ground state of a dimer,
 and retaining those atomic states that have above a given probability.
The remaining details of the procedure are postponed to \secref{ben_details},
 but, most importantly, a threshold probability of $10^{-6}$ chooses
 23 states from 696-dimensional Fock spaces of each respective atom;
 11 of these are neutral states, while 4 are cationic, and 8 are anionic.
The probability threshold is relatively arbitrary, chosen only to provide an accuracy for the dimer
 that is beyond reproach, while still demonstrating the efficiency of the method for larger systems.
This basis, extracted from a dimer at 4.5 {\AA}
\new{
 (the minimum along the 6-31G FCI bonding potential),
}
 was used for all Be fragments,
 regardless of their number or the distance between them.

With the model space for each fragment fixed, the corresponding excitonic Hamiltonian
 is also defined, in principle, per the discussion of fragment-decomposed electronic systems provided in the companion article.
In practice, however, we may apply further approximations.
Already mentioned among these is the neglect of the trimer and tetramer terms in the XR2-CCSD Hamiltonian.
We should also carry along some indication of the scheme by which the fragment states are determined,
 but we suppress this here, since the Hamiltonian is effectively converged in this respect (verified by the results),
 and a systematic taxonomy has not been developed.
Yet another approximation will be made here by assuming that the dimer coupling elements for any pair of fragments
 are independent of the presence of other fragments.
\new{
 This is to say that the effects of Pauli repulsions from neighbors on the dimer interaction are neglected.
 For the weakly overlapping systems here, we expect this to be reasonable.
The Pauli repulsions within
  the pairwise dimer interactions are preserved.
}
\reworked{
We augment the method name here as XR2$'$-CCSD to reflect this approximation.
The reason for doing this is only to simplify the programming tasks necessary to complete this proof-of-principle demonstration.
Although intuitive (and exact for two-fragment systems), this effectively relaxes global antisymmetry.
}
The precise details of the Hamiltonian construction are collected in \secref{ben_details}.

For all calculations, both XR2$'$-CCSD and conventional [henceforth, just CCSD or CCSD(T)],
 the amplitude iterations converged
 when the total energy changed by less than $10^{-12}$ $E_\text{h}$ between iterations.
This is an observation, rather than a criterion, though thresholds in each code were adjusted to arrange for this agreement.
The \textsc{Psi4} 1.0.0 package \cite{Parrish:2017:Psi4} was used
 to obtain both energies and timings for conventional CC variants,
 and also to to run FCI benchmark curves.
The excitonic Hamiltonians for electronic systems were built using one- and two-electron integrals obtained from the
 \textsc{PyQuante} 1.6.5 package \cite{Muller:2014:PyQuante165}, which consists of easily modified Python modules.
The faster, explicit version of the amplitude equations was implemented for these tests (see \appref{code}).
With the further use of a threshold on the Hamiltonian matrix elements
 (requiring only a few human programming hours),
 this computer code was also sped up dramatically, so much that it even lowered the scaling, while preserving results to within machine precision.
Further discussion of this takes place in \secref{ben_details}.
It is worth mentioning that \textsc{Psi4} similarly applies a threshold to the one- and two-electron integrals with which it operates.
Our purpose here, however, is only to provide evidence that XR-CC is a promising approach 
 in relation to a 
 state-of-the-art standard,
 not to yet compete against any specific implementation or input deck.

\subsubsection{Results}

\begin{figure}
  \centering
  \includegraphics[width=14.5cm]{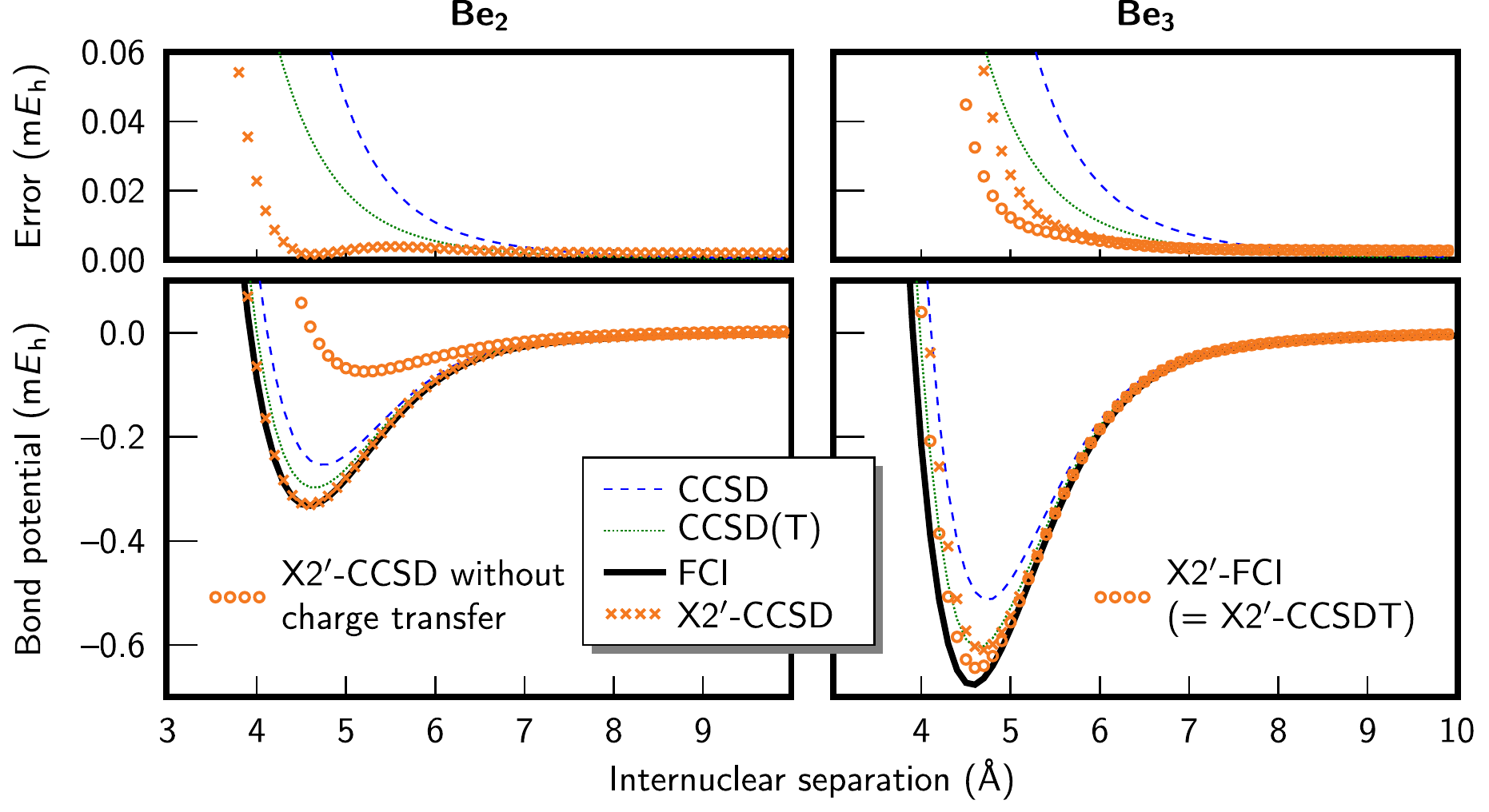}
  \caption{\label{PEcurves}
   The accuracy of XR2$'$-CCSD
   \new{
    in the neighborhood of the binding minima
   }
    is comparable to FCI and CCSD(T) for the dimer and symmetric linear trimer, respectively,
   \new{
    The upper panels give the error for each binding potential (on an expanded scale), relative to FCI.
   }
   Note that the inclusion of charge-transfer fluctuations (\ie{different charge states for the fragments}) is important
    for recovering the interaction.
   By way of analyzing sources of error (Hamiltonian vs.~wavefunction), a complete wavefunction was tested for the
    approximate trimer Hamiltonian, accounting for roughly half of the discrepancy against FCI.
   \reworked{
     XR2$'$-CCSD is nearly exact by construction for the dimer at 4.5 {\AA} (the distance for which the chosen states have been optimized).
   }
  }
\end{figure}


\Figref{PEcurves} shows dissociation curves for the dimer and symmetric linear trimer.
The nearly exact agreement with of XR2$'$-CCSD and FCI
\new{
 in the neighborhood of the minimum
}
 for the dimer is essentially by construction.
Firstly, the wavefunction is complete for dimers.
Secondly, the restriction to dimer terms in the Hamiltonian does not constitute an approximation,
 and the subsequent implied relaxation of global antisymmetry is still exact for dimers.
Thirdly and finally, the single-fragment basis was chosen to
 render the excitonic effective Hamiltonian numerically converged for
 the ground-state dimer
 at 4.5 {\AA}.
For reference, CCSD and CCSD(T) curves show that, in spite of the small system size, this nearly exact agreement is not to be taken for granted.
\new{
 One does notice a finite asymptote to the XR2$'$-CCSD error as the atoms are separated, which has a value of $2.07 \times 10^{-7}$ \textit{E}$_\text{h}$,
  due to the monomer subspaces having been optimized for the dimer interaction (though the residual to restore the monomer ground states could easily be added);
 the errors for conventional CCSD and CCSD(T) go smoothly to zero for this simple system, since it separates into two two-electron problems.
}

One of the most interesting aspects of the dimer plot is the XR2$'$-CCSD calculation that omits anionic and cationic
 states, effectively suppressing charge transfer.
It is primarily interesting that such a model is so easily defined, and it hints at the possible utility 
 of XR-CC approaches to provide insight into bonding character (and basis-set superposition error),
\new{
 similar to existing energy decomposition analyses \cite{Mo:2000:EnergyDecompAnal,Khaliullin:2007:ALMO}.
}
It also validates our earlier claim that charge transfer is important for these test systems 
 (at least, in this basis),
 and it provides a numerical demonstration that XR-CC is equipped to handle such problems.

The trimer curve in \figref{PEcurves} shows that the accuracy does diminish when the dimer-restricted
 Hamiltonian and wavefunction are applied to a trimer system.
\new{
 (For reference the XR2$'$-CCSD error asymptote is $2.73 \times 10^{-7}$ \textit{E}$_\text{h}$.)
}
There are four possible sources of error here.
The first possibility is the truncated wavefunction.
To test this, we would like to run an XR2$'$-CCSDT calculation;
 however, in lieu of implementing the amplitude equations for this, we perform an equivalent XR2$'$-FCI calculation.
The excitonic Hamiltonian is used to populate a Hamiltonian matrix
 in terms of the hypothetical biorthogonal super-system basis states (which need not be constructed, only indexed).
Indeed, we see from \figref{PEcurves} that a substantial fraction of the error is owed to missing trimer correlations.

\begin{figure}
  \centering
  \includegraphics[width=7.5cm]{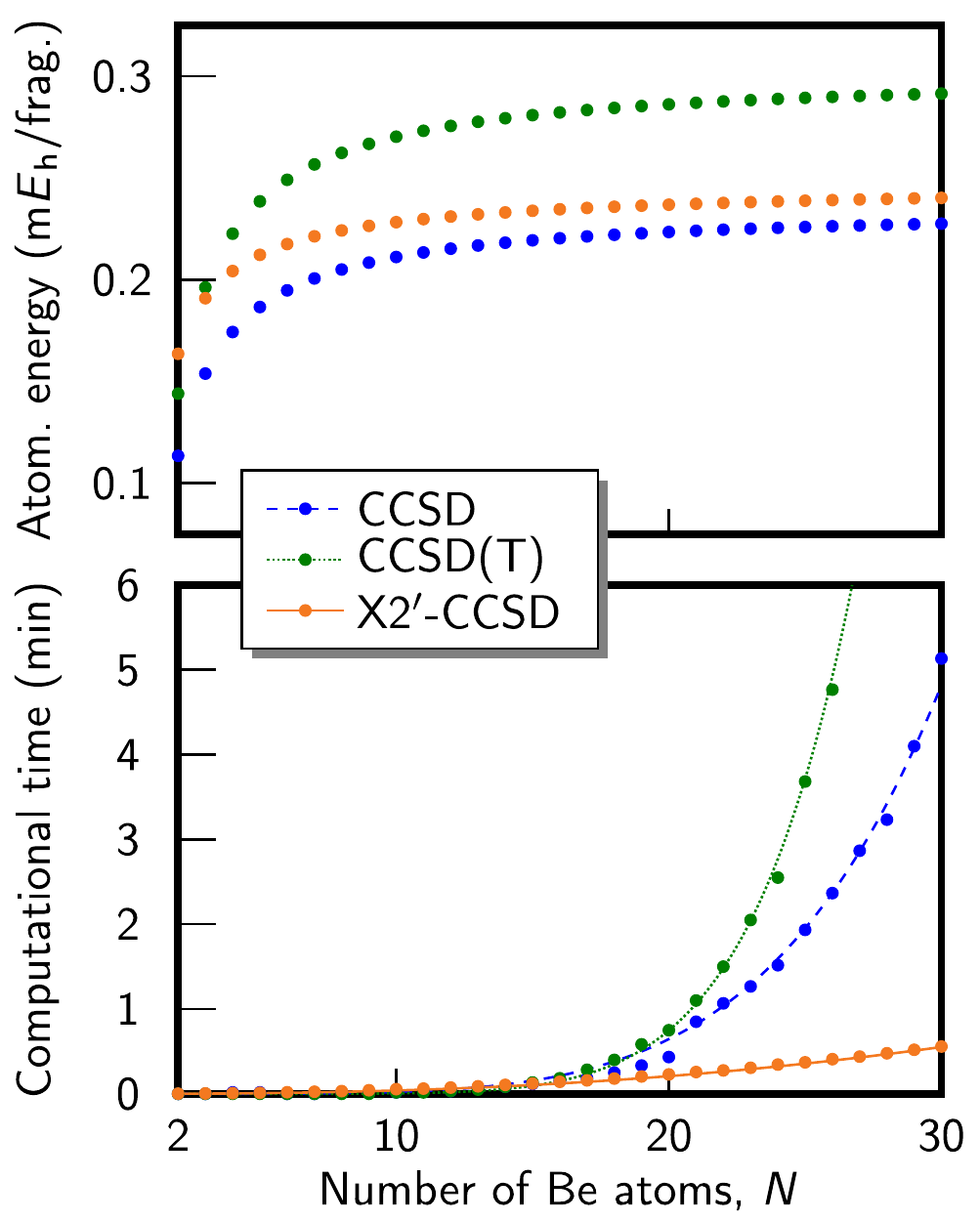}
  \caption{\label{tradeoff}
   The atomization energy (per fragment) and computational effort for each method are plotted for linear chains of Be atoms.
   The accuracy of XR2$'$-CCSD is comparable to CCSD for the larger systems, but the computational effort is much smaller.
   The connecting curves are best-fit monomials; these scale as ${\sim}N^{2.40}$, ${\sim}N^{4.96}$, and ${\sim}N^{7.19}$,
    for XR2$'$-CCSD, CCSD, and (T), respectively.
   For CCSD(T), only the time for the perturbative triples (T) module is plotted.
  }
\end{figure}

The remaining possible errors in the trimer calculations stem from approximations applied to the Hamiltonian.
Foremostly, the Hamiltonian is missing trimer terms, and we have made further approximations
 that allow for implicit global antisymmetry violations.
Only the future implementation of the fully antisymmetric excitonic Hamiltonian will be able to discern the relative sizes of these two errors.
The former seems likely to be substantial, since the electrostatic effects on one atom from a nearby disjoint charge
 transfer are neglected.
Theoretically, there is also a fourth possibility that the single-fragment states optimized for nearest-neighbor interactions
 are not appropriate for capturing next-nearest neighbor interactions.
This would be unexpected, since the dimer dissociation curve was so accurate over such a large range,
 but there could be a non-trivial end-to-end interaction mediated by the middle atom \cite{Dutoi:2010:Ne3,Ambrosetti:2016:WavelikeVdW}.
To test, we lowered the threshold probability for the inclusion of atomic states based on the dimer interaction to $10^{-7}$.
This lowers the dissociation minimum by only 0.8 \textmu$E_\text{h}$,
 so this appears not to contribute significantly to the error.

Since XR2$'$-CCSD agrees with FCI for the dimer but only roughly CCSD(T) for the trimer, the next natural question is whether the
 quality continues to degrade with number of fragments.
To answer this question, the atomization energies of chains of up to 30 atoms (4.5 {\AA} separation) are plotted in \figref{tradeoff},
 along with the same quantity computed using CCSD and CCSD(T).
The error does level off when reckoned per fragment, and the quality always remains a little better than for CCSD.


\begin{table}
  \begin{center}
    \begin{tabular}{r|r}
      \hline
       $N$ & time (min) \\
      \hline
        10 &  0.1 \\
        20 &  0.2 \\
        30 &  0.6 \\
        40 &  1.1 \\
        50 &  1.9 \\
        60 &  2.8 \\
        70 &  4.2 \\
        80 &  5.7 \\
        90 &  7.8 \\
       100 & 10.6 \\
      \hline
    \end{tabular}
  \end{center}
  \caption{\label{Iwin}
    Single-core computational times are given for the XR2$'$-CCSD algorithm for linear chains of $N$ Be atoms.
    The time for assembling the Hamiltonian (a quadratically scaling step) is not included.
  }
\end{table}

Having verified that the XR2$'$-CCSD method is at least as good as CCSD in terms of absolute accuracy for any
 given system, we turn to computational cost.
The run times for systems of up to 30 atoms are plotted in \figref{tradeoff}.
XR2$'$-CCSD is not only faster already for modest $N$, but it has a lower scaling.
The best-fit monomials have scalings of
 $\sim$$N^{2.40}$, $\sim$$N^{4.96}$, and $\sim$$N^{7.19}$ for XR2$'$-CCSD, CCSD, and the perturbative triples module (T), respectively
 (log--log least-squares regression for the largest 10 systems);
\new{
 the expected formal scalings are $\sim$$N^3$, $\sim$$N^6$ and $\sim$$N^7$, respectively.
}
The scaling of XR2$'$-CCSD is below the formally expected behavior due to the thresholding of Hamiltonian matrix elements.
\new{
The conventional CCSD code timings also grow more slowly than expected,
 either because of similar dynamic thresholding or the fact that computation is still dominated by a larger number of lower-scaling terms.}
Whereas the exponent estimates are still rather sensitive to the range chosen for CCSD and (T),
 indicating more general polynomial behavior, XR2$'$-CCSD (generally, XR2-CCSD) has largely reached its asymptotic scaling.
This was verified by fitting to timings for up to 100 atoms, obtaining $\sim$$N^{2.29}$ over this range.
Timings for these calculations are given in \tabref{Iwin}.
Notably, a calculation involving 200 active electrons in 800 spatial orbitals
 has been completed in just 10.6 min,
\reworked{
 giving an atomization energy comparable to that expected from CCSD
}
\new{
 (based on \figref{tradeoff}, since this calculation is not easily completed).}

\new{
It needs to be mentioned that the time for assembling the Hamiltonian is not included here.
The most effective way to optimize single-fragment states and build a Hamiltonian is an area of intense present development;
 the procedure has not been optimized and it exists now as a fragmented set of Python scripts (which are very slow, relative to comparable compiled code).
However, in the companion article we have shown that the Hamiltonian build is both asymptotically quadratically scaling and trivially parallelizable.
To this end, we wish to emphasize that what we have shown here is that the globally coupled part, which is not so trivially parallelized and is generally expected to be more difficult,
 has promising cost and scaling characteristics already on a single core.
}

\subsubsection{Procedural details \label{ben_details}}

Regarding the choice of fragment states,
 an FCI calculation is first performed on a dimer at 4.5 {\AA},
 using a primitive in-house program that extracts the lowest eigenvector by the Lanczos algorithm \cite{Paige:1972:LanczosEigen},
 avoiding construction of the full matrix.
This ground state is then resolved in terms of the set of antisymmetrized tensor products of all 
 16 cationic, 120 neutral, and 560 anionic states for each atom in this basis.
The 1820 possible dianionic states (requiring the transfer of both active electrons from one 
 atom to the other) were excluded, since we expect their contribution to be negligible;
 this is validated by the quality of the results.
The states that we used for this resolution were the precomputed energy eigenstates of the atoms
 (anionic states were trapped by the basis).
A preliminary step resolved the eigenstates of each atom in terms of the dimer molecular orbitals,
 such that antisymmetrization of the tensor products simply requires the Pauli exclusion principle.
The tensor-product basis of atomic eigenstates is not orthonormal,
 but the coefficients nevertheless reflect the relative importance of individual atomic states.

To simplify interpretation, we normalize the aforementioned atomically decomposed coefficient vector (as opposed to the state it represents).
By interpreting the normalized vector as if it referred to an orthonormal tensor-product basis,
 a Fock-space density matrix is constructed for
 both the ``left'' and the ``right'' atom, denoted $\tnsgrk{\rho}_\text{L}$ and $\tnsgrk{\rho}_\text{R}$, each having unit trace.
In order to avoid any left--right bias of the fragment states these density matrices were averaged as
  $\tnsgrk{\rho} = (\tnsgrk{\rho}_\text{L} + \tnsgrk{\rho}_\text{R})/2$.
The eigenvectors of $\tnsgrk{\rho}$ are now the coefficients of a set of fragment states $\{\ket{\psi_{i_m}}\}$,
 where $m$ is any one of the identical fragments.
Each member of this basis is accompanied by an associated probability eigenvalue $\rho_{i_m}$.
Our criterion for constructing the truncated XR-CC fragment basis is to simply
 select the subset of $\{\ket{\psi_{i_m}}\}$ 
 whose members have $\rho_{i_m}$$>$$10^{-6}$ (or $10^{-7}$ to check convergence, as described).

Our brute force procedure for identifying the most important single-fragment states leaves us with dimer coupling information already at hand,
 which is convenient for the construction of the ``XR2$'$'' excitonic Hamiltonian.
Using the resultant $696$$\times$$23$ monomer transformation matrix,
 we project the implicit ($696$$\cdot$$696$)$\times$($696$$\cdot$$696$) dimer Hamiltonian matrix in the non-orthonormal full tensor-product basis
 as an explicit ($23$$\cdot$$23$)$\times$($23$$\cdot$$23$) matrix, where the basis is again not orthonormal.
Interestingly, while this projection is block-diagonal by electron number, it does contain blocks for 2 through 6 active electrons.
These are all necessary;
 although each fluctuation in $\opr{T}$ conserves \textit{overall} charge, these fluctuations interact, and atoms that may have
 changed charge state in their interaction with one atom can interact with yet another.
Per the discussion in the companion article, antisymmetry in the XR-CC formalism
 is accounted for by representing the Hamiltonian matrix elements in a biorthogonal basis.
This is accomplished here for the dimer terms by multiplication of the raw Hamiltonian projection by the inverse of the 
 ($23$$\cdot$$23$)$\times$($23$$\cdot$$23$) overlap matrix for the tensor-product basis of the retained fragment states.
The $23$$\times$$23$ Hamiltonian for an isolated fragment is trivially obtained by projecting
 the original $696$$\times$$696$ fragment Hamiltonian.
In order to isolate the dimer \textit{coupling} elements,
 the tensor product of the single-fragment Hamiltonian matrix with itself (to represent a second identical fragment)
 is subtracted from the dimer Hamiltonian matrix.
Subtractive steps such as this are unnecessary in the procedure given in the companion article.

Hamiltonian matrix elements smaller than $10^{-16}$ $E_\text{h}$ were discarded
 in the XR-CC calculations.
This is done by setting elements below threshold to explicit floating-point zero before the wavefunction iterations.
To the maximum extent possible,
 the tensor contractions were arranged to have the loop over the Hamiltonian indices on the outside of any nesting.
The code could then be made more efficient by testing on the fly whether each Hamiltonian element is identically zero.
The sparsity of the excitonic Hamiltonian for large systems more than compensates for this extra test.
Since optimization is still largely a work in progress, we postpone a thorough discussion to a future publication,
 except to report that, for thresholds in the regime of $10^{-12}$ to $10^{-16}$ $E_\text{h}$,
 the XR2$'$-CCSD energy is unchanged to within machine precision, and the timing curves do not differ substantially.
This hints that the majority of discarded elements are actually very much smaller than these thresholds.
For the record, the \textsc{Psi4} calculations neglected molecular integrals smaller than $10^{-12}$ $E_\text{h}$.

		\section{Conclusion}

In this article, we have laid out the theory and prospective advantages of basing a CC
 wavefunction on fluctuations of fragments between internally correlated states.
Two numerical demonstrations of the promise of this approach have been provided.

As respects application to more chemically interesting systems, it should be pointed out that,
 in some ways, the simulations performed here were already demanding.
The entire interaction between two Be atoms in the basis employed is on the order of 1 kJ/mol, the often-stated
 threshold for chemical accuracy,
 and our absolute errors were small percentages of this near the binding minima.
Furthermore, with such a small excitation energy (2.7 eV, as compared to 7.5 eV for, say, water \cite{Kramida:1997:BeAtomLevels,Mota:2005:WaterElecSpec}),
 the Be atom is quite polarizable (38 $e^2 a_0^2 / E_\text{h}$, as compared to 10 $e^2 a_0^2 / E_\text{h}$ for water \cite{Sahoo:2008:BePolarizability,Murphy:1977:WaterPolarizability}).
The only variables, aside from system size, that determine the
 computational cost of a given XR-CC variant are the number of states per fragment and the sparsity of the couplings in that basis.
At present, it is difficult to know if, for example, coupled water molecules in a high-quality one-electron basis will require much greater
 than 23 many-electron states per fragment.
In case only 23 states were needed, however, then the XR2-CCSD cost
 for 100 water molecules would be the same as for the 100-Be-atom system,
 namely $\sim$11 min (presuming similar Hamiltonian sparsity);
 the extrapolation to 1000 such fragments is 1.5 days.

The timings here are for a computer program that is not yet very sophisticated, or even parallelized.
The fact that simple matrix-element thresholding lowered the observed computational scaling
 speaks well of the \textit{a priori} use of locality in the XR-CC model.
Similar to conventional local-correlation methods
\reworked{
\cite{
 Forner:1985:LocalCorrelation,
 Stoll:1992:LocalCorrDiamond,
 Saebo:1993:LocalCorrelation,
 Schutz:2001:DomainCCSD,
 Maslen:2005:MP4TRIM,
 Subotnik:2006:SmoothLocalCCSD,
 Li:2010:ClusterInMolecule,
 Hattig:2012:PNOMollerPlesset,
 Kristensen:2012:DivideExpandConsolidate,
 Liakos:2015:PNOCClimits},
}
 hard-coded assumptions about the way interactions behave over large distances can further improve efficiency.

An important aspect of the XR-CC model is that it is theoretically
 independent of the level of electronic structure theory used to compute the fragment states.
The most powerful local correlations are already resolved with the introduction of the super-system basis.
Since XR-CC is not tied to any specific level of theory, it is not inherently subject to the shortcomings of any given base method.
The level of theory used for individual fragments could even be multi-reference in nature,
 effectively inverting the traditional paradigm of introducing dynamic correlation into multi-reference
 problems \cite{Mukherjee:1975:OpenShellCC,
 Jeziorski:1981:MRCC,Paldus:1999:CCReview,
 Kinoshita:2005:MRCCtailoredCI,
 Yanai:2006:CanonicalTransMRCC,
 Evangelista:2012:sqicCC}.
The treatment of a super-system also need not be spatially homogeneous;
 fewer states of
 lower quality may be used farther from a region of interest.
Yet, the rigor of the formalism renders all of this systematically improvable.
Naturally,
 the flexibility to have the entire quantum mechanical system subject to an external potential
 (\eg{molecular mechanics embedding}) is still available.

The excited-state regime may also be accessed via
 the established equation-of-motion formalism for linear response \cite{Koch:1990:LRCC,Stanton:1993:EOMCC,Shen:2009:BCCCexcitedstates,Seidler:2011:VCCResponseTheory}.
The ability to straightforwardly proceed from ground-state to excited-state calculations, on account of having a global wavefunction,
 is an important distinction, in contrast to incremental methods \cite{Nolan:2010:Hierarchical,Muller:2011:Incremental,Zimmerman:2017:iFCI},
 which are formally exact fragment decompositions of the ground-state energy and properties only.
\new{
Along with this, it is worthwhile to mention that the standard CC Lagrangian approach to system properties \cite{Helgaker:2002:PurpleBook}
 is also readily applied to XR-CC methods.
The relevant operator would need to be resolved in the excitonic basis, and the procedure for this would be identical
 to that for the Hamiltonian, as outlined in the companion article.
The standard machinery for gradients of the CC energy with respect to nuclear motion is also applicable to XR-CC,
 since the derivative of the effective Hamiltonian with respect to nuclear positions is straightforward;
 it requires only that derivatives of the integrals to be computed,
 since the tacit assumption here is that the single-fragment sub-spaces are fixed, as we have done here
 (they are already presumed adequate to represent interactions with all other species at all ranges).
}

Work in the immediate future will apply the XR-CC algorithm to the more
 carefully implemented excitonic Hamiltonians detailed in the companion article,
 including exact treatment of global (as opposed to pairwise) antisymmetry.
In that article, it was shown that, to within an arbitrary threshold, the number of
 terms in the exact excitonic Hamiltonian 
 scale asymptotically quadratically with system size.
There will admittedly be cases where the cost of constructing this Hamiltonian is prohibitive,
 but also a number of important cases where it is not.
More tractable schemes of determining a preliminary single-fragment basis also need to be developed,
 though it is important to note the continued availability of algorithmic optimizations of the single-fragment bases.
To be concrete, the XR-CC energy is differentiable with respect to fragment-basis rotations,
 and optimization can still be done in advance for dimers only, similar to what was done here.

	\section*{Acknowledgements}

The authors gratefully acknowledge start-up support from the Hornage Fund at the University of the Pacific,
 as well as equipment and travel support provided by the Dean of the College of the Pacific.
The following colleagues are recognized for useful insights during the development of this work:
Arindam Chakraborty,
Gregory J.~O.~Beran,
Oriol Vendrell,
Andreas Dreuw,
Joshua Schrier.

	\section*{Notes}

Supplemental Information available: Alternate derivation of amplitude equations.

\clearpage
{\LARGE \textbf{Appendix}}
\appendix

		\section{Amplitude Equations \label{amplitudes}}

For the XR2-CCSD model, the amplitudes of \eqnref{omega_abstract} evaluate to
 \begin{eqnarray}\label{final_equations}
   \omega_0 &=& 
       \mathcal{H}^{o}_{\dot{x}} t_1^{\dot{x}}
     + \mathcal{H}^{o o}_{\dot{x} \dot{x}'} t_\text{II}^{\dot{x} \dot{x}'}
   \n
   \omega_1^{u_m} &=&
       \mathcal{H}^{u_m}_{o_m} 
     + \mathcal{H}^{o_m}_{o_m} t_1^{u_m}
     + \mathcal{H}^{u_m}_{\dot{x}_m} t_1^{\dot{x}_m}
     + \mathcal{H}^{o}_{\dot{x}} t_2^{\dot{x} u_{m}} 
     - \mathcal{H}^{o_m}_{\dot{x}_m} t_1^{\dot{x}_m} t_1^{u_m}
     + \mathcal{H}^{u_{m} o}_{o_{m} \dot{x}} t_1^{\dot{x}}
     + \mathcal{H}^{o_{m} o}_{o_{m} \dot{x}} t_\text{II}^{\dot{x} u_{m}}
     + \mathcal{H}^{u_{m} o}_{\dot{x}_{m} \dot{x}} t_\text{II}^{\dot{x}_{m} \dot{x}}
       \n &~&
     + 2 \mathcal{H}^{o o}_{\dot{x} \dot{x}'} t_1^{\dot{x}'} t_2^{\dot{x} u_{m}} 
     - 2 \mathcal{H}^{o o_{m}}_{\dot{x} \dot{x}_{m}} t_1^{\dot{x}_{m}} t_2^{\dot{x} u_{m}} 
     - 2 \mathcal{H}^{o_{m} o}_{\dot{x}_{m} \dot{x}} t_\text{II}^{\dot{x}_m \dot{x}} t_1^{u_{m}}
   \n
   \omega_2^{u_{m_1} u_{m_2}} &=&
       \mathcal{H}^{o_{m_1}}_{o_{m_1}} t_2^{u_{m_1} u_{m_2}}
     + \mathcal{H}^{u_{m_1}}_{\dot{x}_{m_1}} t_2^{\dot{x}_{m_1} u_{m_2}}
     - \mathcal{H}^{o_{m_1}}_{\dot{x}_{m_1}} t_1^{\dot{x}_{m_1}} t_2^{u_{m_1} u_{m_2}}
     - \mathcal{H}^{o_{m_2}}_{\dot{x}_{m_2}} t_2^{\dot{x}_{m_2} u_{m_1}} t_1^{u_{m_2}}
       \n &~&
     + \mathcal{H}^{u_{m_1} u_{m_2}}_{o_{m_1} o_{m_2}}
     + \mathcal{H}^{u_{m_1} o_{m_2}}_{o_{m_1} o_{m_2}} t_1^{u_{m_2}}
     + \mathcal{H}^{u_{m_1} u_{m_2}}_{o_{m_1} \dot{x}_{m_2}} t_1^{\dot{x}_{m_2}}
     + \mathcal{H}^{u_{m_1} o}_{o_{m_1} \dot{x}} t_2^{\dot{x} u_{m_2}}
     - \mathcal{H}^{u_{m_1} o_{m_2}}_{o_{m_1} \dot{x}_{m_2}} t_1^{\dot{x}_{m_2}} t_1^{u_{m_2}}
       \n &~&
     + \mathcal{H}^{o_{m_1} o_{m_2}}_{o_{m_1} o_{m_2}} t_\text{II}^{u_{m_1} u_{m_2}}
     + \mathcal{H}^{o_{m_2} u_{m_1}}_{o_{m_2} \dot{x}_{m_1}} t_\text{II}^{\dot{x}_{m_1} u_{m_2}}
     + \mathcal{H}^{o_{m_2} o}_{o_{m_2} \dot{x}} t_2^{\dot{x} u_{m_1}} t_1^{u_{m_2}} 
     - \mathcal{H}^{o_{m_1} o_{m_2}}_{o_{m_1} \dot{x}_{m_2}} t_2^{\dot{x}_{m_2} u_{m_1}} t_1^{u_{m_2}} 
       \n &~&
     + \mathcal{H}^{o_{m_1} o}_{o_{m_1} \dot{x}} t_1^{\dot{x}} t_2^{u_{m_1} u_{m_2}} 
     - \mathcal{H}^{o_{m_1} o_{m_2}}_{o_{m_1} \dot{x}_{m_2}} t_1^{\dot{x}_{m_2}} t_\text{II}^{u_{m_1} u_{m_2}}
     + \mathcal{H}^{u_{m_1} u_{m_2}}_{\dot{x}_{m_1} \dot{x}_{m_2}} t_\text{II}^{\dot{x}_{m_1} \dot{x}_{m_2}}
     + \mathcal{H}^{u_{m_1} o}_{\dot{x}_{m_1} \dot{x}} t_1^{\dot{x}} t_2^{\dot{x}_{m_1} u_{m_2}} 
       \n &~&
     + \mathcal{H}^{u_{m_1} o}_{\dot{x}_{m_1} \dot{x}} t_1^{\dot{x}_{m_1}} t_2^{\dot{x} u_{m_2}} 
     - \mathcal{H}^{u_{m_1} o_{m_2}}_{\dot{x}_{m_1} \dot{x}_{m_2}} t_1^{\dot{x}_{m_2}} t_2^{\dot{x}_{m_1} u_{m_2}} 
     - \mathcal{H}^{u_{m_1} o_{m_2}}_{\dot{x}_{m_1} \dot{x}_{m_2}} t_\text{II}^{\dot{x}_{m_1} \dot{x}_{m_2}} t_1^{u_{m_2}}
     + \mathcal{H}^{o o}_{\dot{x} \dot{x}'} t_2^{\dot{x}' u_{m_1}} t_2^{\dot{x} u_{m_2}}
       \n &~&
     - 2 \mathcal{H}^{o o_{m_2}}_{\dot{x} \dot{x}_{m_2}} t_1^{\dot{x}_{m_2}} t_2^{\dot{x} u_{m_1}} t_1^{u_{m_2}} 
     - 2 \mathcal{H}^{o_{m_2} o}_{\dot{x}_{m_2} \dot{x}} t_1^{\dot{x}} t_2^{\dot{x}_{m_2} u_{m_1}} t_1^{u_{m_2}} 
     + 2 \mathcal{H}^{o_{m_2} o_{m_1}}_{\dot{x}_{m_2} \dot{x}_{m_1}} t_1^{\dot{x}_{m_1}} t_2^{\dot{x}_{m_2} u_{m_1}} t_1^{u_{m_2}} 
       \n &~&
     - 2 \mathcal{H}^{o o_{m_2}}_{\dot{x} \dot{x}_{m_2}} t_2^{\dot{x}_{m_2} u_{m_1}} t_2^{\dot{x} u_{m_2}}
     + \mathcal{H}^{o_{m_1} o_{m_2}}_{\dot{x}_{m_1} \dot{x}_{m_2}} t_2^{\dot{x}_{m_2} u_{m_1}} t_2^{\dot{x}_{m_1} u_{m_2}}
       \n &~&
     - 2 \mathcal{H}^{o_{m_1} o}_{\dot{x}_{m_1} \dot{x}} t_\text{II}^{\dot{x}_{m_1} \dot{x}} t_2^{u_{m_1} u_{m_2}}
     + \mathcal{H}^{o_{m_1} o_{m_2}}_{\dot{x}_{m_1} \dot{x}_{m_2}} t_\text{II}^{\dot{x}_{m_1} \dot{x}_{m_2}} t_\text{II}^{u_{m_1} u_{m_2}}
 \end{eqnarray}
The contraction notation here is essentially the Einstein summation convention, except that contracted indices are explicitly indicated
 by the placement of a dot above them (to distinguish summations from diagonal elements).
If a contracted index is not subscripted by the label of a fragment, summation over all fragments is additionally implied.
We have also made use of the definitions
 \begin{eqnarray}
  t_\text{II}^{u_{m_1} u_{m_2}} = t_2^{u_{m_1} u_{m_2}} ~+~ t_1^{u_{m_1}} t_1^{u_{m_2}}
 \end{eqnarray}
 and
 \begin{eqnarray}\label{matrix_elements}
  \mathcal{H}_0 &=& \sum_{m'} H^{o_{m'}}_{o_{m'}} ~+~ \frac{1}{2} \sum_{m',m''} H^{o_{m'} o_{m''}}_{o_{m'} o_{m''}} \n
  \mathcal{H}^{u_m}_{o_m} &=& H^{u_m}_{o_m} ~+ \sum_{m'} H^{u_m o_{m'}}_{o_m o_{m'}} \n
  \mathcal{H}^{o_m}_{o_m} &=& -\Big(H^{o_m}_{o_m} ~+~ \sum_{m'} H^{o_m o_{m'}}_{o_m o_{m'}}\Big) \n
  \mathcal{H}^{u_m}_{v_m} &=& H^{u_m}_{v_m} ~+ \sum_{m'} H^{u_m o_{m'}}_{v_m o_{m'}} \n
  \mathcal{H}^{o_m}_{u_m} &=& H^{o_m}_{u_m} ~+ \sum_{m'} H^{o_m o_{m'}}_{u_m o_{m'}} \n
  \mathcal{H}^{u_{m_1} u_{m_2}}_{o_{m_1} o_{m_2}} &=& \frac{1}{2} H^{u_{m_1} u_{m_2}}_{o_{m_1} o_{m_2}} \n
  \mathcal{H}^{u_{m_1} o_{m_2}}_{o_{m_1} o_{m_2}} &=& -H^{u_{m_1} o_{m_2}}_{o_{m_1} o_{m_2}} \n
  \mathcal{H}^{u_{m_1} u_{m_2}}_{o_{m_1} v_{m_2}} &=&  H^{u_{m_1} u_{m_2}}_{o_{m_1} v_{m_2}} \n
  \mathcal{H}^{u_{m_1} o_{m_2}}_{o_{m_1} u_{m_2}} &=&  H^{u_{m_1} o_{m_2}}_{o_{m_1} u_{m_2}} \n
  \mathcal{H}^{o_{m_1} o_{m_2}}_{o_{m_1} o_{m_2}} &=&  \frac{1}{2} H^{o_{m_1} o_{m_2}}_{o_{m_1} o_{m_2}} \n
  \mathcal{H}^{o_{m_1} u_{m_2}}_{o_{m_1} v_{m_2}} &=& -H^{o_{m_1} u_{m_2}}_{o_{m_1} v_{m_2}} \n
  \mathcal{H}^{o_{m_1} o_{m_2}}_{o_{m_1} u_{m_2}} &=& -H^{o_{m_1} o_{m_2}}_{o_{m_1} u_{m_2}} \n
  \mathcal{H}^{u_{m_1} u_{m_2}}_{v_{m_1} v_{m_2}} &=& \frac{1}{2} H^{u_{m_1} u_{m_2}}_{v_{m_1} v_{m_2}} \n
  \mathcal{H}^{u_{m_1} o_{m_2}}_{v_{m_1} u_{m_2}} &=&  H^{u_{m_1} o_{m_2}}_{v_{m_1} u_{m_2}} \n
  \mathcal{H}^{o_{m_1} o_{m_2}}_{u_{m_1} u_{m_2}} &=& \frac{1}{2} H^{o_{m_1} o_{m_2}}_{u_{m_1} u_{m_2}} 
 \end{eqnarray}
The matrix elements on the right-hand sides of \eqnref{matrix_elements} are those from the 
 Hamiltonian expansion in the main text of the article,
 except that they are there only defined for $m_1$$<$$m_2$.

The expressions in \eqnref{final_equations} are valid only if the tensors of both amplitudes and Hamiltonian matrix elements
 are symmetric with respect to permutation of indices.
(Undefined elements in the main text are identical to defined elements with permuted indices, where upper and
 lower indices must be simultaneously permuted for Hamiltonian elements, and zero results when $m_1$$=$$m_2$).
For compactness, the the tensor of $\omega_2^{u_{m_1} u_{m_2}}$ values
 does not have such symmetry as it is written in \eqnref{final_equations}.
This expression should be symmetrized (averaged with its transpose) in a \textit{post hoc} step,
 in order that the resulting amplitude update preserves the aforementioned permutational symmetry.

		\section{Code Validation \label{code}}

Two different implementations of the amplitude equations were used,
 in order to boot-strap our development.
A more abstract implementation is based on direct representation of the   
 operators in eqs.~(13) -- (15) of the Supplementary Information, and looping over all terms in the Hamiltonian
 (with some minor adjustments to avoid implicit loops of spuriously high scaling).
The advantage of this approach is that the code contains relatively few lines, which are easily validated
 by eye against the original equations.
This code is quite slow, however, relative to the implementation of the explicit tensor contractions of \eqnref{final_equations}.
The faster implementation based on \eqnref{final_equations} is also implemented with some further factorizations,
 in order to decrease redundancy, and this requires a scratch space for intermediates that scales linearly with the system size.
These aspects are considered details that are
 best discussed in a future publication on optimization (if necessary, given the isomorphism to VCC).

Mutual numerical consistency within a validation chain of independent programs was used to verify that the amplitude equations were implemented correctly.
This relies on a formal/informational mapping of excitonic problems (with distinguishable coordinates) to fermionic ones,
 in which each state of each fragment is a represented by a hypothetical orbital,
 but only transitions within disjoint orbital subsets (corresponding to distinct fragments) are allowed.
In this way, the conventional electronic CC algorithm can be used to run XR-CC calculations
 (albeit inefficiently), using an appropriately mapped integrals tensor.
Of passing interest is that mapping orbitals in fermionic problems
 to distinguishable two-state systems (occupied or not) would theoretically allow
 conventional electronic CC to be run as XR-CC or VCC calculations;
 CCSD would correspond to VCC[4] with four-mode couplings, which has the expected $N^6$ scaling \cite{Seidler:2009:VCCeqnsAutomatic},
 strongly suggesting that algorithmic complexity is conserved across such mappings.

Since it would be arduous to adapt the CC codes of established quantum chemistry packages to
 perform excitonic calculations via the previously mentioned fermionic mapping,
 we implemented the well-known conventional CCSD equations \cite{Helgaker:2002:PurpleBook}, in an in-house program.
The first link of the validation chain was then to check the in-house conventional CCSD code (operating with the usual one- and two-electron integrals)
 against established packages for small electronic systems (\eg{frozen core Be dimer}).
These checks agreed to all meaningful digits (accounting for thresholds).
This served to validate not only our implementation of the conventional CCSD amplitude equations, but also 
 the quasi-Newton/DIIS nonlinear optimization algorithm in which they are embedded.
(All in-house codes use the same CC iteration machinery, aside from the amplitude equations,
 with the internal structure of $\opr{\mathcal{H}}$ and $\opr{T}$ hidden from the generic nonlinear optimizer.)

With our conventional CCSD code validated, it was then fed the matrix elements for small
 oscillator-model fragment Hamiltonians, using the stated fermionic mapping.
Mutual consistency of this fermionically mapped code
 with the aforementioned abstract implementation of the XR2-CCSD amplitude equations then 
 served to validate both that the mapping was done correctly and that the abstract XR2-CCSD implementation was correct.
The Be-fragment Hamiltonians for small numbers of fragments were then formatted for both this slower abstract implementation
 and the faster explicit implementation, allowing us to thereby validate the more efficient implementation.
For all of the comparisons between our three completely independent implementations of the XR2-CCSD amplitude equations
 (fermionically mapped, abstract, and explicit),
 test results always agreed to within machine precision.

\pagebreak
  
\setcounter{equation}{0}
\setcounter{figure}{0}
\setcounter{table}{0}
\setcounter{page}{1}
\makeatletter
\renewcommand{\theequation}{S\arabic{equation}}
\renewcommand{\thefigure}{S\arabic{figure}}
\renewcommand{\bibnumfmt}[1]{[S#1]}
\renewcommand{\citenumfont}[1]{S#1}

\begin{center}
  \textbf{Supplementary Information: Derivation of Amplitude Equations for Two-index Fluctuation Operators}
\end{center}

We begin by defining operators that have properties closely related to the normal ordering of field operators
 \begin{eqnarray}
  \opr{e}^{o_m}_{u_m} &=&   \opr{\tau}^{o_m}_{u_m} \n
  \opr{f}^{o_m}_{o_m} &=& 1-\opr{\tau}^{o_m}_{o_m} \n
  \opr{f}^{v_m}_{u_m} &=&   \opr{\tau}^{v_m}_{u_m} \n
  \opr{d}^{u_m}_{o_m} &=&   \opr{\tau}^{u_m}_{o_m}
 \end{eqnarray}
Any operator originally expressed in terms of the set of fluctuation operators $\{\opr{\tau}^{j}_{i}\}$ may
 be expressed in terms of these new operators by simple substitution.
The indices $u_m$, $v_m$, $w_m$, $x_m$ refer to any state of fragment $m$ except the reference state, which is denoted $o_m$.
These letters have been chosen to be reminiscent of the ``occupied/zeroth'' and ``unoccupied/virtual'' nomenclature familiar from conventional
 many-body theory, and yet be distinct from its usual notation.
Similarly, the letters used for the four different types of operators stand for ``excitation,'' ``flat,'' and ``de-excitation.''

Of primary importance is that
 all such operators, except excitations, produce the null state when acting on the many-fragment reference.
In general, it will be valuable to consider a generalized definition of normal ordering for strings of such operators,
 in which any excitation operators appear to the left of all other operators.
In this way, any normal-ordered string will destroy the reference state unless it contains only excitations (or is a scalar).
For $\opr{\mathcal{H}}$, this is trivial to accomplish;
 since no string therein contains more than one operator that acts on any given fragment,
 all operators in these strings commute.
The substitution of $\opr{\tau}^{o_m}_{o_m}$ by $1-\opr{f}^{o_m}_{o_m}$ does introduce a constant into
 the normal-ordered Hamiltonian, however, which is equal to $\bra{\Psi^O}\opr{\mathcal{H}}\ket{\Psi_O}$,
 in analogy to the conventional normal-ordered Hamiltonian in terms of field operators.
$\ket{\Psi^O}$ is the biorthogonal complement of the reference state $\ket{\Psi_O}$ (see the companion of the main article for complete discussion).

The cluster operator $\opr{T}$, which must be repeatedly commuted with $\opr{\mathcal{H}}$,
 is composed only of operators of the excitation class.
The following special cases of the commutator in eq.~(3) of the Main Text will therefore be useful
 \begin{eqnarray}\label{special_comm}
   {}[\opr{e}^{o_m}_{u_m}, \opr{e}^{o_m}_{w_m}] &=& 0 \n
   {}[\opr{f}^{o_m}_{o_m}, \opr{e}^{o_m}_{w_m}] &=& \opr{e}^{o_m}_{w_m} \n
   {}[\opr{f}^{v_m}_{u_m}, \opr{e}^{o_m}_{w_m}] &=& \delta_{v_m w_m} \opr{e}^{o_m}_{u_m} \n
   {}[\opr{d}^{u_m}_{o_m}, \opr{e}^{o_m}_{w_m}] &=& \delta_{u_m w_m} - (\delta_{u_m w_m}\opr{f}^{o_m}_{o_m} + \opr{f}^{u_m}_{w_m})
 \end{eqnarray}
 where we recall that commutators between operators belonging to different fragments are always zero.
This now makes explicit earlier arguments concerning truncation of the BCH expansion,
 in that repeated commutations of any operator with excitations always ends with zero.
No operator survives more than two such nested commutations.

In order to be generic at some points, we will let $\opr{g}_m$ represent an arbitrary
 single-fragment operator on fragment $m$.
This could be any one of $\opr{e}^{o_m}_{u_m}$, $\opr{f}^{o_m}_{o_m}$, $\opr{f}^{v_m}_{u_m}$, or $\opr{d}^{u_m}_{o_m}$,
 or linear combinations thereof, and we permit this notation to also include linear combinations that contain constants.
We then furthermore define the abbreviations
 \begin{eqnarray}\label{comm_abbrev}
   \opr{g}_m^{[w_m]}     &=& [\opr{g}_m, \opr{e}^{o_m}_{w_m}] \n
   \opr{g}_m^{[w_m][x_m]} &=& [[\opr{g}_m, \opr{e}^{o_m}_{w_m}],\opr{e}^{o_m}_{x_m}]
 \end{eqnarray}
 noting that these are themselves single-fragment operators (for purposes of recursion).
According to the forgoing observation, all nested commutations higher than second order are zero.
Using this notation, combined with recursion of the well-known formula for commutation with a simple operator product,
 we then resolve commutation of a single-fragment operator with a string of $M$ excitations as
 \begin{eqnarray}\label{commute_w_string}
  [\opr{g}_m, \opr{e}^{o_{m_1}}_{w_{m_1}}\cdots\opr{e}^{o_{m_M}}_{w_{m_M}}] &=&
  \sum_{i=1}^{M} \delta_{m,m_i} ~ \opr{e}^{o_{m_1}}_{w_{m_1}}\cdots\opr{e}^{o_{m_{(i-1)}}}_{w_{m_{(i-1)}}}\opr{e}^{o_{m_{(i+1)}}}_{w_{m_{(i+1)}}} \cdots\opr{e}^{o_{m_M}}_{w_{m_M}}\,\opr{g}_m^{[w_{m_i}]} 
 \end{eqnarray}
 under the condition that the indices $m_1, \cdots m_M$ identify distinct fragments,
 thus allowing rearrangement of the operator strings.
Also, under this restriction, maximally one term of the right-hand side of \eqnref{commute_w_string} is nonzero.

The $M$th-order part of the cluster operator can be written as
 \begin{eqnarray}
  \opr{T}_M &=& \sum_{m_1} \sum_{m_2>m_1} \cdots \sum_{m_M>m_{M-1}} \bigg(\sum_{w_{m_1}}\cdots\sum_{w_{m_M}} t_M^{w_{m_1}\cdots w_{m_M}} \opr{e}^{o_{m_1}}_{w_{m_1}}\cdots\opr{e}^{o_{m_M}}_{w_{m_M}} \bigg) \n
            &=& \frac{1}{M!}\sum_{m_1} \cdots \sum_{m_M} \bigg(\sum_{w_{m_1}}\cdots\sum_{w_{m_M}} t_M^{w_{m_1}\cdots w_{m_M}} \opr{e}^{o_{m_1}}_{w_{m_1}}\cdots\opr{e}^{o_{m_M}}_{w_{m_M}} \bigg) 
 \end{eqnarray}
 where the total cluster operator $\opr{T} = \sum_{M} \opr{T}_M$ is a summation over all orders $M$$\geq$$1$ present in the Ansatz.
In the first line, we sum over all unique $M$-tuples of fragments,
 and in the second line we account for redundancy by dividing by $M!$ and insisting that the tensor of amplitudes $\textbf{\textit{t}}_M$ 
 (containing elements $t_M^{w_{m_1}\cdots w_{m_M}}$) is invariant with respect to all index permutations.
We may also insist that an amplitude is zero if any two indices belong to the same fragment, effectively removing from further consideration
 those operator strings in which the same fragment occurs twice (though this is not strictly necessary, since such strings themselves contribute zero).
Using \eqnref{commute_w_string}, we then arrive at
 \begin{eqnarray}
  [\opr{g}_m, \opr{T}_M] &=& \frac{1}{M!}\sum_{m_1} \cdots \sum_{m_M} \bigg( \sum_{w_{m_1}}\cdots\sum_{w_{m_M}} t_M^{w_{m_1}\cdots w_{m_M}} [\opr{g}_m, \opr{e}^{o_{m_1}}_{w_{m_1}}\cdots\opr{e}^{o_{m_M}}_{w_{m_M}}] \bigg) \n
                         &=& \frac{1}{(M-1)!} \sum_{m_1\neq m} \cdots \sum_{m_{M-1}\neq m} \bigg( \sum_{w_{m_1}}\cdots\sum_{w_{m_{M-1}}}\sum_{w_m} t_M^{w_{m_1}\cdots w_{m_{M-1}} w_m} \, \opr{e}^{o_{m_1}}_{w_{m_1}}\cdots\opr{e}^{o_{m_{M-1}}}_{w_{m_{M-1}}} \, \opr{g}_m^{[w_m]} \bigg) \n
                         &=& \sum_{w_m} \opr{T}_{M-1}^{w_m} \, \opr{g}_m^{[w_m]} \n
   \opr{T}_M^{w_m} &=& \frac{1}{M!} \sum_{m_1\neq m} \cdots \sum_{m_M\neq m} \bigg( \sum_{w_{m_1}}\cdots\sum_{w_{m_M}} t_{M+1}^{w_{m_1}\cdots w_{m_M} w_m} \, \opr{e}^{o_{m_1}}_{w_{m_1}}\cdots\opr{e}^{o_{m_M}}_{w_{m_M}} \bigg) \n
                   &=& \frac{1}{M!} \sum_{m_1}       \cdots \sum_{m_M}       \bigg( \sum_{w_{m_1}}\cdots\sum_{w_{m_M}} t_{M+1}^{w_{m_1}\cdots w_{m_M} w_m} \, \opr{e}^{o_{m_1}}_{w_{m_1}}\cdots\opr{e}^{o_{m_M}}_{w_{m_M}} \bigg) \n
  \opr{T}_0^{w_m} &=& t_1^{w_m}
 \end{eqnarray}
The logical process by which this is deduced is to decompose each of the $M$ summations over all fragments in the first line into one component for fragment $m$ and a summation over all
 fragments other than $m$.
This gives a total of $2^M$ terms when the resulting $M$-fold binomial product is expanded.
Only $M$ of these terms survive, since the relevant amplitude is zero if any two indices both belong to fragment $m$ and the commutator is zero if no fragment index
 is equal to $m$; therefore, only one fragment can be equal to $m$ in any surviving term, of which there are $M$ choices.
Each surviving term contains $M-1$ summations over the other fragments.
Since all of the other fragments and indices are summed over, and since all of the operators in the strings commute and amplitudes are permutationally symmetric,
 these $M$ terms are all identical, simply reducing the prefactor to $1/(M-1)!$ for the single such term written explicitly in the second line.
In the last line, the restrictions on the summations are removed since the amplitudes of the superfluous terms thereby introduced are zero.
This has the advantage of giving $\opr{T}_M^{w_m}$ an identical structure to $\opr{T}_M$ (for $M$$\neq$$0$), but with different amplitudes,
 permitting us to use recursion to immediately write
 \begin{eqnarray}\label{commute_normal_order}
  [\opr{g}_m, \opr{T}_M^{w_{m'}}] &=& \sum_{w_m} \opr{T}_{M-1}^{w_m w_{m'}} \, \opr{g}_{m}^{[w_m]} \n
   \opr{T}_M^{w_m w_{m'}} &=& \frac{1}{M!} \sum_{m_1} \cdots \sum_{m_M} \bigg( \sum_{w_{m_1}}\cdots\sum_{w_{m_M}} t_{M+2}^{w_{m_1}\cdots w_{m_M} w_m w_{m'}} \, \opr{e}^{o_{m_1}}_{w_{m_1}}\cdots\opr{e}^{o_{m_M}}_{w_{m_M}} \bigg) \n
  \opr{T}_0^{w_m w_{m'}} &=& t_2^{w_m w_{m'}}
 \end{eqnarray}
\Eqnref{commute_normal_order} will be necessary later
 to bring the transformation of products of operators into generalized normal-ordered form.

It will now be expedient to define the following summations over $M$, in parallel to the definition of $\opr{T}$ itself
 \begin{eqnarray}
  \opr{T}^{w_m} &=& \sum_{M} \opr{T}^{w_m}_{M-1} \n
  \opr{T}^{w_m w_{m'}} &=& \sum_{M} \opr{T}^{w_m w_{m'}}_{M-2}
 \end{eqnarray}
 where the summation over $M$ is over all orders originally in the user Ansatz.
(For consistency, both $\opr{T}_M^{w_m}$ and $\opr{T}_M^{w_m w_{m'}}$ are defined as zero for $M$$<$$0$.)
These allow us to write more compactly
 \begin{eqnarray}
   {}[\opr{g}_m, \opr{T}] &=& \sum_{w_m} \opr{T}^{w_m} \, \opr{g}_m^{[w_m]} \n
   {}[\opr{g}_m, \opr{T}^{w_{m'}}] &=& \sum_{w_m} \opr{T}^{w_m w_{m'}} \, \opr{g}_m^{[w_m]}
 \end{eqnarray}
By recursion, we then also arrive at
 \begin{eqnarray}
  {}[[\opr{g}_m, \opr{T}], \opr{T}] &=& \sum_{w_m} \sum_{x_m} \opr{T}^{w_m} \opr{T}^{x_m} \opr{g}_m^{[w_m][x_m]}
 \end{eqnarray}
 where it is clear that any triply nested commutator vanishes.
This allows us to then use the BCH expansion to finally write, for any single-fragment operator,
  \begin{eqnarray}\label{trans_single}
    \bar{g}_m ~=~ \ee^{-\opr{T}}\opr{g}_m\ee^{\opr{T}}
              ~=~ \opr{g}_m + \sum_{w_m} \opr{T}^{w_m} \opr{g}_m^{[w_m]} + \frac{1}{2}\sum_{w_m} \sum_{x_m} \opr{T}^{w_m} \opr{T}^{x_m} \opr{g}_m^{[w_m][x_m]}
  \end{eqnarray}
 where the abbreviation as $\bar{g}_m$ will be convenient later.
Importantly, as a consequence of the restrictions on the indices in $\opr{T}$ referring to distinct fragments in each string,
 each term in this expansion is already in the aforementioned generalized normal-ordered form.

The forgoing suffices to perform the similarity transformation of the single-fragment parts of $\opr{\mathcal{H}}$, and it can also be
 used to construct expressions for the higher-fragment-order terms.
In order to proceed with the latter task, we first note that any operator $\opr{o}$ may be decomposed as $(\opr{o})_\text{x} + (\opr{o})_\text{o}$,
 such that,
 \begin{eqnarray}
  (\opr{o})_\text{x}\ket{\Psi_O} &=& \opr{o}\ket{\Psi_O} \n
  (\opr{o})_\text{o}\ket{\Psi_O} &=& 0
 \end{eqnarray}
Although such a decomposition is not unique as just described, if only the action upon the reference state $\ket{\Psi_O}$ is relevant for a specific purpose,
 then any convenient such partitioning will suffice.
If $\opr{o}$ is already written as a linear combination of normal-ordered strings,
 then the straightforward choice of $(\opr{o})_\text{o}$ consists of summing all such terms whose string contains at least one operator that is not an excitation.
The corresponding choice of $(\opr{o})_\text{x}$ then consists of the remaining terms, \ie{linear combination of strings of excitations only, and perhaps a constant}.
For convenience, we also allow commutator brackets to be subscripted as $[,]_\text{x}$, indicating that only the 
 constant and excitation part of the normal-ordered form of the result are retained.

The central task in a given CC iteration may now be framed in terms of resolving the operator
 $\opr{\Omega} = (\ee^{-\opr{T}}\opr{\mathcal{H}}\ee^{\opr{T}})_\text{x}$.
The constant part of $\opr{\Omega}$ is the CC pseudo-energy, which can also be written as $\prj{\Psi^O}\ket{\Omega}$,
 with $\ket{\Omega} = \opr{\Omega}\ket{\Psi_O}$.
Projections $\prj{\Psi^I}\ket{\Omega}$ 
 onto excited complement configurations of the form $\bra{\Psi^I} = \bra{\Psi^O}\opr{d}^{u_{m_1}}_{o_{m_1}}\cdots\opr{d}^{u_{m_M}}_{o_{m_M}}$
 are used to determine the iterative updates to the associated amplitudes $t_M^{u_{m_1}\cdots u_{m_M}}$,
 in conjunction with multiplication by index-dependent scalars (algorithm-dependent preconditioners).
This involves computing excitations in $\opr{\Omega}$ up to the user-specified Ansatz order.

We will now proceed to similarity transform the individual operator strings found in the Hamiltonian.
For interaction terms up to the maximum possible number of single-fragment operators for electronic systems (four) we have
\begin{eqnarray}\label{abstract_interactions}
 \big(\ee^{-\opr{T}} \opr{g}_m \ee^{\opr{T}}\big)_\text{x} &=& (\bar{g}_m)_\text{x} \n
 \big( \ee^{-\opr{T}} \opr{g}_{m_1} \opr{g}_{m_2} \ee^{\opr{T}} \big)_\text{x}
 &=& \big(\bar{g}_{m_1}\bar{g}_{m_2}\big)_\text{x} \n
 &=& \big( ( (\bar{g}_{m_1})_\text{x} + (\bar{g}_{m_1})_\text{o} ) (\bar{g}_{m_2})_\text{x} \big)_\text{x} \n
 &=& \big(
           (\bar{g}_{m_1})_\text{x}(\bar{g}_{m_2})_\text{x}
          +(\bar{g}_{m_1})_\text{o}(\bar{g}_{m_2})_\text{x}
     \big)_\text{x} \n
 &=&
           (\bar{g}_{m_1})_\text{x}(\bar{g}_{m_2})_\text{x}
          +[(\bar{g}_{m_1})_\text{o},(\bar{g}_{m_2})_\text{x}]_\text{x} \n
 \big( \ee^{-\opr{T}} \opr{g}_{m_1} \opr{g}_{m_2} \opr{g}_{m_3} \ee^{\opr{T}} \big)_\text{x}
 &=&
             (\bar{g}_{m_1})_\text{x}(\bar{g}_{m_2})_\text{x}(\bar{g}_{m_3})_\text{x}
            +(\bar{g}_{m_1})_\text{x}[(\bar{g}_{m_2})_\text{o},(\bar{g}_{m_3})_\text{x}]_\text{x} \n
 &~&        +[(\bar{g}_{m_1})_\text{o},(\bar{g}_{m_2})_\text{x}(\bar{g}_{m_3})_\text{x}]_\text{x} 
            +[(\bar{g}_{m_1})_\text{o},[(\bar{g}_{m_2})_\text{o},(\bar{g}_{m_3})_\text{x}]_\text{x}]_\text{x} \n
 \big(\ee^{-\opr{T}} \opr{g}_{m_1} \opr{g}_{m_2} \opr{g}_{m_3} \opr{g}_{m_4}\ee^{\opr{T}} \big)_\text{x}
 &=&
             (\bar{g}_{m_1})_\text{x}(\bar{g}_{m_2})_\text{x}(\bar{g}_{m_3})_\text{x}(\bar{g}_{m_4})_\text{x}
            +(\bar{g}_{m_1})_\text{x}(\bar{g}_{m_2})_\text{x}[(\bar{g}_{m_3})_\text{o},(\bar{g}_{m_4})_\text{x}]_\text{x} \n
 &~&        +(\bar{g}_{m_1})_\text{x}[(\bar{g}_{m_2})_\text{o},(\bar{g}_{m_3})_\text{x}(\bar{g}_{m_4})_\text{x}]_\text{x}
            +(\bar{g}_{m_1})_\text{x}[(\bar{g}_{m_2})_\text{o},[(\bar{g}_{m_3})_\text{o},(\bar{g}_{m_4})_\text{x}]_\text{x}]_\text{x} \n
 &~&        +[(\bar{g}_{m_1})_\text{o}, (\bar{g}_{m_2})_\text{x}(\bar{g}_{m_3})_\text{x}(\bar{g}_{m_4})_\text{x}]_\text{x}
            +[(\bar{g}_{m_1})_\text{o}, (\bar{g}_{m_2})_\text{x}[(\bar{g}_{m_3})_\text{o},(\bar{g}_{m_4})_\text{x}]_\text{x}]_\text{x} \n
 &~&        +[(\bar{g}_{m_1})_\text{o}, [(\bar{g}_{m_2})_\text{o},(\bar{g}_{m_3})_\text{x}(\bar{g}_{m_4})_\text{x}]_\text{x}]_\text{x}
            +[(\bar{g}_{m_1})_\text{o}, [(\bar{g}_{m_2})_\text{o},[(\bar{g}_{m_3})_\text{o},(\bar{g}_{m_4})_\text{x}]_\text{x}]_\text{x}]_\text{x} \n 
 \end{eqnarray}
The results for trimers and tetramers are obtained by recurring the procedure shown for the dimer term.
The logic in resolving the dimer term is as follows:
 after inserting $1=\ee^{\opr{T}}\ee^{-\opr{T}}$ between the two single-fragment operators, each of the resulting normal-ordered
 transformed operators is divided into the parts that do and do not destroy the reference.
Only the $()_\text{x}$ part of the right-most operator needs to be retained, since inclusion of the $()_\text{o}$ part
 simply results in additional normal-ordered terms that all destroy the reference, and therefore do not survive
 the outermost retention of only non-reference-destroying terms.
Likewise, inclusion of the second term of the commutator shown does not change anything, since it consists only of
 reference-destroying terms that are not retained.
However, expressing the result in terms of this commutator will prove valuable;
 since the second argument to the commutator consists only of constants and excitation strings, it is clear that fragment rank is hereby reduced.

To proceed, we now require explicit forms of the transformed single-fragment operators.
Inserting the results of \eqnref{special_comm} into \eqnsref{comm_abbrev}{trans_single},
 and again using an overbar to denote a similarity transformed operator, we have
 \begin{eqnarray}\label{normalmonomers}
  \bar{e}^{o_m}_{u_m}  &=& \opr{e}^{o_m}_{u_m} \n
  \bar{f}^{o_m}_{o_m}  &=& \opr{f}^{o_m}_{o_m} + \sum_{w_m} \opr{T}^{w_m} \opr{e}^{o_m}_{w_m} \n
  \bar{f}^{v_m}_{u_m}  &=& \opr{f}^{v_m}_{u_m} + \opr{T}^{v_m} \opr{e}^{o_m}_{u_m} \n
  \bar{d}^{u_m}_{o_m}  
    &=& \opr{d}^{u_m}_{o_m} - \big(\opr{T}^{u_m}\opr{f}^{o_m}_{o_m} + \sum_{w_m} \opr{T}^{w_m} \opr{f}^{u_m}_{w_m} \big) + \opr{T}^{u_m}\big(1-\sum_{w_m} \opr{T}^{w_m} \opr{e}^{o_m}_{w_m}\big)
 \end{eqnarray}
 each of which is easily divided into an $()_\text{x}$ part (the last or only term) and other $()_\text{o}$ terms.

We will now confine our attention here to the dimer Hamiltonian terms that were implemented for this work.
The procedure for higher-fragment-order terms is a straightforward repetition of this procedure, albeit, increasingly tedious.
If we now arbitrarily decide that normal ordering also involves having any
 de-excitation operators to the far right and having any virtual-rearrangement flat operators ($\opr{f}^{v_m}_{u_m}$) to the right of any reference-hole-check flat operators ($\opr{f}^{o_m}_{o_m}$),
 then we have 10 classes of dimer terms in the normal-ordered Hamiltonian:
 $\opr{e}^{o_{m_1}}_{u_{m_1}}\opr{e}^{o_{m_2}}_{u_{m_2}}$, 
 $\opr{e}^{o_{m_1}}_{u_{m_1}}\opr{f}^{o_{m_2}}_{o_{m_2}}$, 
 $\opr{e}^{o_{m_1}}_{u_{m_1}}\opr{f}^{v_{m_2}}_{u_{m_2}}$, 
 $\opr{e}^{o_{m_1}}_{u_{m_1}}\opr{d}^{u_{m_2}}_{o_{m_2}}$, 
 $\opr{f}^{o_{m_1}}_{o_{m_1}}\opr{f}^{o_{m_2}}_{o_{m_2}}$, 
 $\opr{f}^{o_{m_1}}_{o_{m_1}}\opr{f}^{v_{m_2}}_{u_{m_2}}$, 
 $\opr{f}^{o_{m_1}}_{o_{m_1}}\opr{d}^{u_{m_2}}_{o_{m_2}}$, 
 $\opr{f}^{v_{m_1}}_{u_{m_1}}\opr{f}^{v_{m_2}}_{u_{m_2}}$, 
 $\opr{f}^{v_{m_1}}_{u_{m_1}}\opr{d}^{u_{m_2}}_{o_{m_2}}$,
 $\opr{d}^{u_{m_1}}_{o_{m_1}}\opr{d}^{u_{m_2}}_{o_{m_2}}$.
The non-zero commutators of \eqnref{abstract_interactions} that are then needed are
 \begin{eqnarray}\label{normaldimers}
  {}[(\bar{f}^{o_{m_1}}_{o_{m_1}})_\text{o},(\bar{f}^{o_{m_2}}_{o_{m_2}})_\text{x}]_\text{x} 
      &=& \sum_{w_{m_1},w_{m_2}} \opr{T}^{w_{m_1},w_{m_2}} \opr{e}^{o_{m_1}}_{w_{m_1}} \opr{e}_{w_{m_2}}^{o_{m_2}} \n
  {}[(\bar{f}^{o_{m_1}}_{o_{m_1}})_\text{o},(\bar{f}^{v_{m_2}}_{u_{m_2}})_\text{x}]_\text{x} 
      &=& \sum_{w_{m_1}} \opr{T}^{w_{m_1},v_{m_2}} \opr{e}^{o_{m_1}}_{w_{m_1}} \opr{e}_{u_{m_2}}^{o_{m_2}} \n
  {}[(\bar{f}^{o_{m_1}}_{o_{m_1}})_\text{o},(\bar{d}^{u_{m_2}}_{o_{m_2}})_\text{x}]_\text{x}
      &=&
          \sum_{w_{m_1}} \opr{T}^{w_{m_1} u_{m_2}} \opr{e}^{o_{m_1}}_{w_{m_1}}
        - \sum_{w_{m_1},w_{m_2}} \Big(\opr{T}^{w_{m_1} u_{m_2}} \opr{T}^{w_{m_2}} + \opr{T}^{w_{m_1} w_{m_2}} \opr{T}^{u_{m_2}}\Big) \opr{e}^{o_{m_1}}_{w_{m_1}} \opr{e}^{o_{m_2}}_{w_{m_2}} \n
  {}[(\bar{f}^{v_{m_1}}_{u_{m_1}})_\text{o},(\bar{f}^{v_{m_2}}_{u_{m_2}})_\text{x}]_\text{x}
      &=& \opr{T}^{v_{m_1} v_{m_2}} \opr{e}^{o_{m_1}}_{u_{m_1}} \opr{e}^{o_{m_2}}_{u_{m_2}} \n
  {}[(\bar{f}^{v_{m_1}}_{u_{m_1}})_\text{o},(\bar{d}^{u_{m_2}}_{o_{m_2}})_\text{x}]_\text{x}
      &=&
          \opr{T}^{v_{m_1} u_{m_2}} \opr{e}^{o_{m_1}}_{u_{m_1}} 
        - \sum_{w_{m_2}} \Big( \opr{T}^{v_{m_1} u_{m_2}} \opr{T}^{w_{m_2}} + \opr{T}^{v_{m_1} w_{m_2}} \opr{T}^{u_{m_2}}\Big) \opr{e}^{o_{m_1}}_{u_{m_1}} \opr{e}^{o_{m_2}}_{w_{m_2}} \n
  {}[(\bar{d}^{u_{m_1}}_{o_{m_1}})_\text{o},(\bar{d}^{u_{m_2}}_{o_{m_2}})_\text{x}]_\text{x}
      &=& 
        \opr{T}^{u_{m_1} u_{m_2}} \n
      &~& - \sum_{w_{m_1}} \Big(\opr{T}^{u_{m_1}} \opr{T}^{w_{m_1} u_{m_2}} + \opr{T}^{w_{m_1}} \opr{T}^{u_{m_1} u_{m_2}} \Big) \opr{e}^{o_{m_1}}_{w_{m_1}} \n
      &~& - \sum_{w_{m_2}} \Big(\opr{T}^{u_{m_1} u_{m_2}} \opr{T}^{w_{m_2}} + \opr{T}^{u_{m_1} w_{m_2}} \opr{T}^{u_{m_2}} \Big) \opr{e}^{o_{m_2}}_{w_{m_2}} \n
      &~& + \sum_{w_{m_1}, w_{m_2}} \Big(~~~~~ \opr{T}^{u_{m_1} u_{m_2}} \opr{T}^{w_{m_1} w_{m_2}} + \opr{T}^{u_{m_1} w_{m_2}} \opr{T}^{w_{m_1} u_{m_2}}  \n
      &~& ~~~~~~~~~~~~~~~ +~\opr{T}^{u_{m_1}} \opr{T}^{w_{m_1} u_{m_2}} \opr{T}^{w_{m_2}} + \opr{T}^{w_{m_1}} \opr{T}^{u_{m_1} u_{m_2}} \opr{T}^{w_{m_2}} \n
      &~& ~~~~~~~~~~~~~~~ +~\opr{T}^{u_{m_1}} \opr{T}^{w_{m_1} w_{m_2}} \opr{T}^{u_{m_2}} + \opr{T}^{w_{m_1}} \opr{T}^{u_{m_1} w_{m_2}} \opr{T}^{u_{m_2}} \Big) \opr{e}^{o_{m_1}}_{w_{m_1}} \opr{e}^{o_{m_2}}_{w_{m_2}} \n
 \end{eqnarray}
 which make use of the fact that $m_1\neq m_2$.

We now further confine our attention to the XR2-CCSD model, for which we also have
 \begin{eqnarray}\label{x2ccsdT}
   \opr{T}^{w_m} &=& t_1^{w_m} + \sum_{m', w_{m'}} t_2^{w_m w_{m'}} \opr{e}^{o_{m'}}_{w_{m'}} \n
   \opr{T}^{w_{m_1} w_{m_2}} &=& t_2^{w_{m_1} w_{m_2}}
 \end{eqnarray}
In order to then resolve $\opr{\Omega}$ for this model, \eqnsref{normaldimers}{x2ccsdT} are inserted into \eqnref{abstract_interactions} for each class of dimer terms,
 and the resulting similarity transformed
 operators are linearly combined according to the matrix elements of the normal-ordered Hamiltonian,
 such that if we decompose $\opr{\Omega}$ as
 \begin{eqnarray}\label{omega_def}
  \opr{\Omega} ~=~ \omega_0 ~+~ \sum_{m,u_m} \omega_1^{u_m} \opr{e}^{o_m}_{u_m} ~+ \frac{1}{2} \sum_{m_1,u_{m_1}} \sum_{m_2,u_{m_2}} \omega_2^{u_{m_1} u_{m_2}} \opr{e}^{o_{m_1}}_{u_{m_1}} \opr{e}^{o_{m_2}}_{u_{m_2}} ~+~ \cdots
 \end{eqnarray}
 the coefficients for up to dimer terms can be written as
 %
 %
 \begin{eqnarray}
   \omega_0 &=& 
       \mathcal{H}^{o}_{\dot{x}} t_1^{\dot{x}}
     + \mathcal{H}^{o o}_{\dot{x} \dot{x}'} t_\text{II}^{\dot{x} \dot{x}'}
   \n
   \omega_1^{u_m} &=&
       \mathcal{H}^{u_m}_{o_m} 
     + \mathcal{H}^{o_m}_{o_m} t_1^{u_m}
     + \mathcal{H}^{u_m}_{\dot{x}_m} t_1^{\dot{x}_m}
     + \mathcal{H}^{o}_{\dot{x}} t_2^{\dot{x} u_{m}} 
     - \mathcal{H}^{o_m}_{\dot{x}_m} t_1^{\dot{x}_m} t_1^{u_m}
     + \mathcal{H}^{u_{m} o}_{o_{m} \dot{x}} t_1^{\dot{x}}
     + \mathcal{H}^{o_{m} o}_{o_{m} \dot{x}} t_\text{II}^{\dot{x} u_{m}}
     + \mathcal{H}^{u_{m} o}_{\dot{x}_{m} \dot{x}} t_\text{II}^{\dot{x}_{m} \dot{x}}
       \n &~&
     + 2 \mathcal{H}^{o o}_{\dot{x} \dot{x}'} t_1^{\dot{x}'} t_2^{\dot{x} u_{m}} 
     - 2 \mathcal{H}^{o o_{m}}_{\dot{x} \dot{x}_{m}} t_1^{\dot{x}_{m}} t_2^{\dot{x} u_{m}} 
     - 2 \mathcal{H}^{o_{m} o}_{\dot{x}_{m} \dot{x}} t_\text{II}^{\dot{x}_m \dot{x}} t_1^{u_{m}}
   \n
   \omega_2^{u_{m_1} u_{m_2}} &=&
       \mathcal{H}^{o_{m_1}}_{o_{m_1}} t_2^{u_{m_1} u_{m_2}}
     + \mathcal{H}^{u_{m_1}}_{\dot{x}_{m_1}} t_2^{\dot{x}_{m_1} u_{m_2}}
     - \mathcal{H}^{o_{m_1}}_{\dot{x}_{m_1}} t_1^{\dot{x}_{m_1}} t_2^{u_{m_1} u_{m_2}}
     - \mathcal{H}^{o_{m_2}}_{\dot{x}_{m_2}} t_2^{\dot{x}_{m_2} u_{m_1}} t_1^{u_{m_2}}
       \n &~&
     + \mathcal{H}^{u_{m_1} u_{m_2}}_{o_{m_1} o_{m_2}}
     + \mathcal{H}^{u_{m_1} o_{m_2}}_{o_{m_1} o_{m_2}} t_1^{u_{m_2}}
     + \mathcal{H}^{u_{m_1} u_{m_2}}_{o_{m_1} \dot{x}_{m_2}} t_1^{\dot{x}_{m_2}}
     + \mathcal{H}^{u_{m_1} o}_{o_{m_1} \dot{x}} t_2^{\dot{x} u_{m_2}}
     - \mathcal{H}^{u_{m_1} o_{m_2}}_{o_{m_1} \dot{x}_{m_2}} t_1^{\dot{x}_{m_2}} t_1^{u_{m_2}}
       \n &~&
     + \mathcal{H}^{o_{m_1} o_{m_2}}_{o_{m_1} o_{m_2}} t_\text{II}^{u_{m_1} u_{m_2}}
     + \mathcal{H}^{o_{m_2} u_{m_1}}_{o_{m_2} \dot{x}_{m_1}} t_\text{II}^{\dot{x}_{m_1} u_{m_2}}
     + \mathcal{H}^{o_{m_2} o}_{o_{m_2} \dot{x}} t_2^{\dot{x} u_{m_1}} t_1^{u_{m_2}} 
     - \mathcal{H}^{o_{m_1} o_{m_2}}_{o_{m_1} \dot{x}_{m_2}} t_2^{\dot{x}_{m_2} u_{m_1}} t_1^{u_{m_2}} 
       \n &~&
     + \mathcal{H}^{o_{m_1} o}_{o_{m_1} \dot{x}} t_1^{\dot{x}} t_2^{u_{m_1} u_{m_2}} 
     - \mathcal{H}^{o_{m_1} o_{m_2}}_{o_{m_1} \dot{x}_{m_2}} t_1^{\dot{x}_{m_2}} t_\text{II}^{u_{m_1} u_{m_2}}
     + \mathcal{H}^{u_{m_1} u_{m_2}}_{\dot{x}_{m_1} \dot{x}_{m_2}} t_\text{II}^{\dot{x}_{m_1} \dot{x}_{m_2}}
     + \mathcal{H}^{u_{m_1} o}_{\dot{x}_{m_1} \dot{x}} t_1^{\dot{x}} t_2^{\dot{x}_{m_1} u_{m_2}} 
       \n &~&
     + \mathcal{H}^{u_{m_1} o}_{\dot{x}_{m_1} \dot{x}} t_1^{\dot{x}_{m_1}} t_2^{\dot{x} u_{m_2}} 
     - \mathcal{H}^{u_{m_1} o_{m_2}}_{\dot{x}_{m_1} \dot{x}_{m_2}} t_1^{\dot{x}_{m_2}} t_2^{\dot{x}_{m_1} u_{m_2}} 
     - \mathcal{H}^{u_{m_1} o_{m_2}}_{\dot{x}_{m_1} \dot{x}_{m_2}} t_\text{II}^{\dot{x}_{m_1} \dot{x}_{m_2}} t_1^{u_{m_2}}
     + \mathcal{H}^{o o}_{\dot{x} \dot{x}'} t_2^{\dot{x}' u_{m_1}} t_2^{\dot{x} u_{m_2}}
       \n &~&
     - 2 \mathcal{H}^{o o_{m_2}}_{\dot{x} \dot{x}_{m_2}} t_1^{\dot{x}_{m_2}} t_2^{\dot{x} u_{m_1}} t_1^{u_{m_2}} 
     - 2 \mathcal{H}^{o_{m_2} o}_{\dot{x}_{m_2} \dot{x}} t_1^{\dot{x}} t_2^{\dot{x}_{m_2} u_{m_1}} t_1^{u_{m_2}} 
     + 2 \mathcal{H}^{o_{m_2} o_{m_1}}_{\dot{x}_{m_2} \dot{x}_{m_1}} t_1^{\dot{x}_{m_1}} t_2^{\dot{x}_{m_2} u_{m_1}} t_1^{u_{m_2}} 
       \n &~&
     - 2 \mathcal{H}^{o o_{m_2}}_{\dot{x} \dot{x}_{m_2}} t_2^{\dot{x}_{m_2} u_{m_1}} t_2^{\dot{x} u_{m_2}}
     + \mathcal{H}^{o_{m_1} o_{m_2}}_{\dot{x}_{m_1} \dot{x}_{m_2}} t_2^{\dot{x}_{m_2} u_{m_1}} t_2^{\dot{x}_{m_1} u_{m_2}}
       \n &~&
     - 2 \mathcal{H}^{o_{m_1} o}_{\dot{x}_{m_1} \dot{x}} t_\text{II}^{\dot{x}_{m_1} \dot{x}} t_2^{u_{m_1} u_{m_2}}
     + \mathcal{H}^{o_{m_1} o_{m_2}}_{\dot{x}_{m_1} \dot{x}_{m_2}} t_\text{II}^{\dot{x}_{m_1} \dot{x}_{m_2}} t_\text{II}^{u_{m_1} u_{m_2}}
 \end{eqnarray}
 where we have defined
 \begin{eqnarray}
  t_\text{II}^{u_{m_1} u_{m_2}} = t_2^{u_{m_1} u_{m_2}} ~+~ t_1^{u_{m_1}} t_1^{u_{m_2}}
 \end{eqnarray}
The contraction notation here is essentially the Einstein summation convention, except that contracted indices are explicitly indicated
 by the placement of a dot above them (to distinguish summations from diagonal elements).
If a contracted index is not subscripted by the label of a fragment, summation over all fragments is additionally implied.
The matrix elements with which the amplitudes are contracted are from the normal-ordered Hamiltonian,
 such that
 \begin{eqnarray}\label{matrix_elements_SI}
  \mathcal{H}_0 &=& \sum_{m'} H^{o_{m'}}_{o_{m'}} ~+~ \frac{1}{2} \sum_{m',m''} H^{o_{m'} o_{m''}}_{o_{m'} o_{m''}} \n
  \mathcal{H}^{u_m}_{o_m} &=& H^{u_m}_{o_m} ~+ \sum_{m'} H^{u_m o_{m'}}_{o_m o_{m'}} \n
  \mathcal{H}^{o_m}_{o_m} &=& -\Big(H^{o_m}_{o_m} ~+~ \sum_{m'} H^{o_m o_{m'}}_{o_m o_{m'}}\Big) \n
  \mathcal{H}^{u_m}_{v_m} &=& H^{u_m}_{v_m} ~+ \sum_{m'} H^{u_m o_{m'}}_{v_m o_{m'}} \n
  \mathcal{H}^{o_m}_{u_m} &=& H^{o_m}_{u_m} ~+ \sum_{m'} H^{o_m o_{m'}}_{u_m o_{m'}} \n
  \mathcal{H}^{u_{m_1} u_{m_2}}_{o_{m_1} o_{m_2}} &=& \frac{1}{2} H^{u_{m_1} u_{m_2}}_{o_{m_1} o_{m_2}} \n
  \mathcal{H}^{u_{m_1} o_{m_2}}_{o_{m_1} o_{m_2}} &=& -H^{u_{m_1} o_{m_2}}_{o_{m_1} o_{m_2}} \n
  \mathcal{H}^{u_{m_1} u_{m_2}}_{o_{m_1} v_{m_2}} &=&  H^{u_{m_1} u_{m_2}}_{o_{m_1} v_{m_2}} \n
  \mathcal{H}^{u_{m_1} o_{m_2}}_{o_{m_1} u_{m_2}} &=&  H^{u_{m_1} o_{m_2}}_{o_{m_1} u_{m_2}} \n
  \mathcal{H}^{o_{m_1} o_{m_2}}_{o_{m_1} o_{m_2}} &=&  \frac{1}{2} H^{o_{m_1} o_{m_2}}_{o_{m_1} o_{m_2}} \n
  \mathcal{H}^{o_{m_1} u_{m_2}}_{o_{m_1} v_{m_2}} &=& -H^{o_{m_1} u_{m_2}}_{o_{m_1} v_{m_2}} \n
  \mathcal{H}^{o_{m_1} o_{m_2}}_{o_{m_1} u_{m_2}} &=& -H^{o_{m_1} o_{m_2}}_{o_{m_1} u_{m_2}} \n
  \mathcal{H}^{u_{m_1} u_{m_2}}_{v_{m_1} v_{m_2}} &=& \frac{1}{2} H^{u_{m_1} u_{m_2}}_{v_{m_1} v_{m_2}} \n
  \mathcal{H}^{u_{m_1} o_{m_2}}_{v_{m_1} u_{m_2}} &=&  H^{u_{m_1} o_{m_2}}_{v_{m_1} u_{m_2}} \n
  \mathcal{H}^{o_{m_1} o_{m_2}}_{u_{m_1} u_{m_2}} &=& \frac{1}{2} H^{o_{m_1} o_{m_2}}_{u_{m_1} u_{m_2}} 
 \end{eqnarray}
 where the raw matrix elements on the right-hand sides of \eqnref{matrix_elements_SI} are those from the 
 Hamiltonian expansion in the main article,
 except that they are there only defined for $m_1$$<$$m_2$.
As with the amplitudes we may simply assert that the otherwise undefined elements are identical
 to that in which the upper and lower indices are simultaneously permuted, with zero resulting for $m_1$$=$$m_2$.
In finalizing the above expressions we have often made use of such permutational symmetries to better organize the indices.
It is worth noting that, as its elements are written, the tensor $\tnsgrk{\omega}_2$ is not symmetric with respect to permutation of indices,
 as this would require the inclusion of additional terms that constitute substantially similar computations.
This tensor can simply be symmetrized without resulting in any change to $\opr{\Omega}$ in \eqnref{omega_def};
 this can be done efficiently in a \textit{post hoc} step.
In practice, this would be required, 
 in order that the resulting update to $\tns{t}_2$ preserves the asserted permutational symmetry,
 on which the validity of the foregoing derivation rests.


\begin{thebibliography}{88}
\providecommand{\url}[1]{\texttt{#1}}
\providecommand{\urlprefix}{URL }

\bibitem{Kitaura:1999:FMO}
K. {Kitaura}, E. {Ikeo}, T. {Asada}, T. {Nakano} and M. {Uebayasi},  Chem.
  Phys. Lett.  \textbf{313}, 701 (1999).

\bibitem{Hirata:2005:ESEmbedding}
S. {Hirata}, M. {Valiev}, M. {Dupuis}, S.S. {Xantheas}, S. {Sugiki} and H.
  {Sekino},  Mol. Phys.  \textbf{103}, 2255 (2005).

\bibitem{Fedorov:2007:FMO}
D.G. {Fedorov} and K. {Kitaura},  J. Phys. Chem.  \textbf{111}, 6904 (2007).

\bibitem{Kamiya:2008:EStaticEmbedding}
M. {Kamiya}, S. {Hirata} and M. {Valiev},  J. Chem. Phys.  \textbf{128}, 074103
  (2008).

\bibitem{Jacobson:2011:XPS}
L.D. {Jacobson} and J.M. {Herbert},  J. Chem. Phys.  \textbf{134}, 094118
  (2011).

\bibitem{Huang:2011:PotFuncEmbedding}
C. {Huang} and E.A. {Carter},  J. Chem. Phys.  \textbf{135}, 194104 (2011).

\bibitem{Manby:2012:ExactDFTEmbedding}
F.R. {Manby}, M. {Stella}, J.D. {Goodpaster} and T.F. {Miller}, III,  J. Chem.
  Theory Comput.  \textbf{8}, 2564 (2012).

\bibitem{Bulik:2014:DFTEmbedding}
I.W. {Bulik}, W. {Chen} and G.E. {Scuseria},  J. Chem. Phys.  \textbf{141},
  054113 (2014).

\bibitem{Dresselhaus:2015:WFembedDFT}
T. {Dresselhaus} and J. {Neugebauer},  Theor. Chem. Acc.  \textbf{134}, 97
  (2015).

\bibitem{Jeziorski:1994:SAPT}
B. {Jeziorski}, R. {Moszynski} and K. {Szalewicz},  Chem. Rev.  \textbf{94},
  1887 (1994).

\bibitem{Byrd:2015:MolClusterCCPT}
J.N. {Byrd}, N. {Jindal}, R.W. {Molt}, Jr., R.J. {Bartlett}, B.A. {Sanders} and
  V.F. {Lotrich},  Mol. Phys.  \textbf{113}, 3459 (2015).

\bibitem{Nolan:2010:Hierarchical}
S.J. {Nolan}, P.J. {Bygrave}, N.L. {Allan} and F.R. {Manby},  J. Phys.:
  Condens. Matter  \textbf{22}, 074201 (2010).

\bibitem{Muller:2011:Incremental}
C. {M\"{u}ller}, D. {Usvyat} and H. {Stoll},  Phys. Rev. B  \textbf{83}, 245136
  (2011).

\bibitem{Beran:2016:CrystalPolymorphRev}
G.J.O. {Beran},  Chem. Rev.  \textbf{116}, 5567 (2016).

\bibitem{Forner:1985:LocalCorrelation}
W. {F\"{o}rner}, J. {Ladik}, P. {Otto} and J. {\v{C}\'{i}\u{z}ek},  Chem. Phys.
   \textbf{97}, 251 (1985).

\bibitem{Stoll:1992:LocalCorrDiamond}
H. {Stoll},  Phys. Rev. B  \textbf{46}, 6700 (1992).

\bibitem{Saebo:1993:LocalCorrelation}
S. {Saeb{\o}} and P. {Pulay},  Annu. Rev. Phys. Chem.  \textbf{44}, 213 (1993).

\bibitem{Schutz:2001:DomainCCSD}
M. {Sch\"{u}tz} and H.J. {Werner},  J. Chem. Phys.  \textbf{114}, 661 (2001).

\bibitem{Maslen:2005:MP4TRIM}
P.E. {Maslen}, A.D. {Dutoi}, M.S. {Lee}, Y. {Shao} and M. {Head-Gordon},  Mol.
  Phys.  \textbf{103}, 425 (2005).

\bibitem{Subotnik:2006:SmoothLocalCCSD}
J.E. {Subotnik}, A. {Sodt} and M. {Head-Gordon},  J. Chem. Phys.  \textbf{125},
  074116 (2006).

\bibitem{Li:2010:ClusterInMolecule}
W. {Li} and P. {Piecuch},  J. Phys. Chem. A  \textbf{114}, 8644 (2010).

\bibitem{Hattig:2012:PNOMollerPlesset}
C. {H\"{a}ttig}, D.P. {Tew} and B. {Helmich},  J. Chem. Phys.  \textbf{136},
  204105 (2012).

\bibitem{Kristensen:2012:DivideExpandConsolidate}
K. {Kristensen}, I.M. {H{\o}yvik}, B. {Jansik}, P. {J{\o}rgensen}, T.
  {Kj{\ae}rgaard}, S. {Reine} and J. {Jakowski},  Phys. Chem. Chem. Phys.
  \textbf{14}, 15706 (2012).

\bibitem{Liakos:2015:PNOCClimits}
D.G. {Liakos}, M. {Sparta}, M.K. {Kesharwani}, J.M.L. {Martin} and F. {Neese},
  J. Chem. Theory Comput.  \textbf{11}, 1525 (2015).

\bibitem{Kotliar:2006:DynamicalMeanField}
G. {Kotliar}, S.Y. {Savrasov}, K. {Haule}, V.S. {Oudovenko}, O. {Parcollet} and
  C.A. {Marianetti},  Rev. Mod. Phys.  \textbf{78}, 865 (2006).

\bibitem{Gordon:2011:FMOReview}
M.S. {Gordon}, D.G. {Fedorov}, S.R. {Pruitt} and L.V. {Slipchenko},  Chem. Rev.
   \textbf{112}, 632 (2011).

\bibitem{Richard:2012:FragUnifiedView}
R.M. {Richard} and J.M. {Herbert},  J. Chem. Phys.  \textbf{137}, 064113
  (2012).

\bibitem{Collins:2015:EnergyBasedFragMethods}
M.A. {Collins} and R.P.A. {Bettens},  Chem. Rev.  \textbf{115}, 5607 (2015).

\bibitem{Raghavachari:2015:FragmentReview}
K. {Raghavachari} and A. {Saha},  Chem. Rev.  \textbf{115}, 5643 (2015).

\bibitem{Huang:2008:ElecStructSolids}
P. {Huang} and E.A. {Carter},  Annu. Rev. Phys. Chem.  \textbf{59}, 261 (2008).

\bibitem{Wesolowski:2015:EmbeddingReview}
T.A. {Wesolowski}, S. {Shedge} and X. {Zhou},  Chem. Rev.  \textbf{115}, 5891
  (2015).

\bibitem{Christiansen:2004:VCCformalism}
O. {Christiansen},  J. Chem. Phys.  \textbf{120}, 2140 (2004).

\bibitem{Seidler:2010:VCCReview}
P. {Seidler} and O. {Christiansen}, in \emph{Challenges and Advances in
  Computational Chemistry and Physics}, edited by P. {C{\'a}rsky}, J. {Paldus}
  and J. {Pittner}, Vol.~11  (Springer Netherlands, Dordrecht, 2010), p. 491.

\bibitem{Faucheaux:2015:HighOrderVCC}
J.A. {Faucheaux} and S. {Hirata},  J. Chem. Phys.  \textbf{143}, 134105 (2015).

\bibitem{Frenkel:1931:Excitons}
J. {Frenkel},  Phys. Rev.  \textbf{37}, 17 (1931).

\bibitem{Davydov:1971:MolecularExcitons}
A.S. {Davydov}, \emph{Theory of Molecular Excitons}   (Plenum, New York, 1971).

\bibitem{May:2011:ChgNrgXfer}
V. {May} and O. {K\"{u}hn}, \emph{Charge and Energy Transfer Dynamics in
  Molecular Systems}   (Wiley, Weinheim, Germany, 2011).

\bibitem{Sisto:2014:AbInitoExciton}
A. {Sisto}, D.R. {Glowacki} and T.J. {Martinez},  Acc. Chem. Res.  \textbf{47},
  2857 (2014).

\bibitem{Morrison:2015:AbInitioExciton}
A.F. {Morrison} and J.M. {Herbert},  J. Phys. Chem. Lett.  \textbf{6}, 4390
  (2015).

\bibitem{Stone:2013:IntermolForces}
A. {Stone}, \emph{The Theory of Intermolecular Forces}   (Oxford University
  Press, Oxford, 2013).

\bibitem{Korona:2009:SAPTCC}
T. {Korona},  J. Chem. Theory Comput.  \textbf{5}, 2663 (2009).

\bibitem{Podeszwa:2007:SAPTDFT3Body}
R. {Podeszwa} and K. {Szalewicz},  J. Chem. Phys.  \textbf{126}, 194101 (2007).

\bibitem{Cui:2006:InductionFourthOrder}
J. {Cui}, H. {Liu} and K.D. {Jordan},  J. Phys. Chem. B  \textbf{110}, 18872
  (2006).

\bibitem{Lao:2016:LargeSystemMBE}
K.U. {Lao}, K.Y. {Liu}, R.M. {Richard} and J.M. {Herbert},  J. Chem. Phys.
  \textbf{144}, 164105 (2016).

\bibitem{Riley:2010:IntermolForceDissect}
K.E. {Riley}, M. {Pito\v{n}\'{a}k}, P. {Jure\v{c}ka} and P. {Hobza},  Chem.
  Rev.  \textbf{110}, 5023 (2010).

\bibitem{Ambrosetti:2016:WavelikeVdW}
A. {Ambrosetti}, N. {Ferri}, R.A. {DiStasio}, Jr. and A. {Tkatchenko},  Science
   \textbf{351}, 1171 (2016).

\bibitem{White:1992:DMRG}
S.R. {White},  Phys. Rev. Lett.  \textbf{69}, 2863 (1992).

\bibitem{Chan:2011:DMRGReview}
G.K.L. {Chan} and S. {Sharma},  Annu. Rev. Phys. Chem.  \textbf{62}, 465
  (2011).

\bibitem{Hedegard:2015:DMRGdispersion}
E.D. {Hedeg{\aa}rd}, S. {Knecht}, J.S. {Kielberg}, H.J.A. {Jensen} and M.
  {Reiher},  J. Chem. Phys.  \textbf{142}, 224108 (2015).

\bibitem{Parker:2014:ActiveSpaceDecompDMRG}
S.M. {Parker} and T. {Shiozaki},  J. Chem. Phys.  \textbf{141}, 211102 (2014).

\bibitem{Kim:2015:ActiveSpaceCovalent}
I. {Kim}, S.M. {Parker} and T. {Shiozaki},  J. Chem. Theory Comput.
  \textbf{11}, 3636 (2015).

\bibitem{Li:2004:BCCCtheory}
S. {Li},  J. Chem. Phys.  \textbf{120}, 5017 (2004).

\bibitem{Fang:2008:BCCCbondbreaking}
T. {Fang}, J. {Shen} and S. {Li},  J. Chem. Phys.  \textbf{128}, 224107 (2008).

\bibitem{Helgaker:2002:PurpleBook}
T. {Helgaker}, P. {Jorgensen} and J. {Olsen}, \emph{Molecular
  Electronic-Structure Theory}   (Wiley, Sussex, 2002).

\bibitem{Christiansen:2004:VCCimplementation}
O. {Christiansen},  J. Chem. Phys.  \textbf{120}, 2149 (2004).

\bibitem{Seidler:2008:VCCefficient}
P. {Seidler}, M.B. {Hansen} and O. {Christiansen},  J. Chem. Phys.
  \textbf{128}, 154113 (2008).

\bibitem{Seidler:2009:VCCeqnsAutomatic}
P. {Seidler} and O. {Christiansen},  J. Chem. Phys.  \textbf{131}, 234109
  (2009).

\bibitem{Ewing:1970:MetalDimers}
R.H. {Ewing} and A.M. {Mellor},  J. Chem. Phys.  \textbf{53}, 2983 (1970).

\bibitem{Blomberg:1978:Be2}
M.R.A. {Blomberg} and P.E.M. {Siegbahn},  Int. J. Quantum Chem.  \textbf{14},
  583 (1978).

\bibitem{Bondybey:1984:TheoryExptBe2}
V.E. {Bondybey} and J.H. {English},  J. Chem. Phys.  \textbf{80}, 568 (1984).

\bibitem{Kowalski:2005:Be3}
K. {Kowalski}, S. {Hirata}, M. {W{\l}och}, P. {Piecuch} and T.L. {Windus},  J.
  Chem. Phys.  \textbf{123}, 074319 (2005).

\bibitem{Merritt:2009:Be2experimental}
J.M. {Merritt}, V.E. {Bondybey} and M.C. {Heaven},  Science  \textbf{324}, 1548
  (2009).

\bibitem{El_Khatib:2014:Be2StaticCorrelation}
M. {El Khatib}, G.L. {Bendazzoli}, S. {Evangelisti}, W. {Helal}, T.
  {Leininger}, L. {Tenti} and C. {Angeli},  J. Phys. Chem. A  \textbf{118},
  6664 (2014).

\bibitem{Sharma:2014:Be2DMRG}
S. {Sharma}, T. {Yanai}, G.H. {Booth}, C.J. {Umrigar} and G.K.L. {Chan},  J.
  Chem. Phys.  \textbf{140}, 104112 (2014).

\bibitem{Meshkov:2014:Be2potential}
V.V. {Meshkov}, A.V. {Stolyarov}, M.C. {Heaven}, C. {Haugen} and R.J. {LeRoy},
  J. Chem. Phys.  \textbf{140}, 064315 (2014).

\bibitem{Dutoi:2004:Radicals}
A.D. {Dutoi}, Y. {Jung} and M. {Head-Gordon},  J. Phys. Chem. A  \textbf{108},
  10270 (2004).

\bibitem{Khaliullin:2009:WaterDimerCT}
R.Z. {Khaliullin}, A.T. {Bell} and M. {Head-Gordon},  Chem. Eur. J.
  \textbf{15}, 851 (2009).

\bibitem{Parrish:2017:Psi4}
R.M. {Parrish}, L.A. {Burns}, D.G.A. {Smith}, A.C. {Simmonett}, A.E.
  {DePrince}, III, E.G. {Hohenstein}, U. {Bozkaya}, A.Y. {Sokolov}, R. {Di
  Remigio}, R.M. {Richard}, J.F. {Gonthier}, A.M. {James}, H.R. {McAlexander},
  A. {Kumar}, M. {Saitow}, X. {Wang}, B.P. {Pritchard}, P. {Verma}, H.F.
  {Schaefer}, III, K. {Patkowski}, R.A. {King}, E.F. {Valeev}, F.A.
  {Evangelista}, J.M. {Turney}, T.D. {Crawford} and C.D. {Sherrill},  J. Chem.
  Theory Comput.  \textbf{13}, 3185 (2017).

\bibitem{Muller:2014:PyQuante165}
R.P. {Muller},  \texttt{http://pyquante.sourceforge.net/}  \textbf{Some bibtex
  formats seem not to be able to handle software references, so I'm pretending
  this is an article.}, 0 (2014, [Accessed 25-October-2016]).

\bibitem{Mo:2000:EnergyDecompAnal}
Y. {Mo}, J. {Gao} and S.D. {Peyerimhoff},  J. Chem. Phys.  \textbf{112}, 5530
  (2000).

\bibitem{Khaliullin:2007:ALMO}
R.Z. {Khaliullin}, E.A. {Cobar}, R.C. {Lochan}, A.T. {Bell} and M.
  {Head-Gordon},  J. Phys. Chem. A  \textbf{111}, 8753 (2007).

\bibitem{Dutoi:2010:Ne3}
A.D. {Dutoi}, L.S. {Cederbaum}, M. {Wormit}, J.H. {Starcke} and A. {Dreuw},  J.
  Chem. Phys.  \textbf{132}, 144302 (2010).

\bibitem{Paige:1972:LanczosEigen}
C.C. {Paige},  IMA J. Appl. Math.  \textbf{10}, 373 (1972).

\bibitem{Kramida:1997:BeAtomLevels}
A. {Kramida} and W.C. {Martin},  J. Phys. Chem. Ref. Data  \textbf{26}, 1185
  (1997).

\bibitem{Mota:2005:WaterElecSpec}
R. {Mota}, R. {Parafita}, A. {Giuliani}, M.J. {Hubin-Franskin}, J.M.C.
  {Louren\c{c}o}, G. {Garcia}, S.V. {Hoffmann}, N.J. {Mason}, P.A. {Ribeiro},
  M. {Raposo} and P. {Lim\~{a}o-Vieira},  Chem. Phys. Lett.  \textbf{416}, 152
  (2005).

\bibitem{Sahoo:2008:BePolarizability}
B.K. {Sahoo} and B.P. {Das},  Phys. Rev. A  \textbf{77}, 062516 (2008).

\bibitem{Murphy:1977:WaterPolarizability}
W.F. {Murphy},  J. Chem. Phys.  \textbf{67}, 5877 (1977).

\bibitem{Mukherjee:1975:OpenShellCC}
D. {Mukherjee}, R.K. {Moitra} and A. {Mukhopadhyay},  Mol. Phys.  \textbf{30},
  1861 (1975).

\bibitem{Jeziorski:1981:MRCC}
B. {Jeziorski} and H.J. {Monkhorst},  Phys. Rev. A  \textbf{24}, 1668 (1981).

\bibitem{Paldus:1999:CCReview}
J. {Paldus} and X. {Li},  Adv. Chem. Phys.  \textbf{110}, 1 (1999).

\bibitem{Kinoshita:2005:MRCCtailoredCI}
T. {Kinoshita}, O. {Hino} and R.J. {Bartlett},  J. Chem. Phys.  \textbf{123},
  074106 (2005).

\bibitem{Yanai:2006:CanonicalTransMRCC}
T. {Yanai} and G.K.L. {Chan},  J. Chem. Phys.  \textbf{124}, 194106 (2006).

\bibitem{Evangelista:2012:sqicCC}
F.A. {Evangelista}, M. {Hanauer}, A. {K\"{o}hn} and J. {Gauss},  J. Chem. Phys.
   \textbf{136}, 204108 (2012).

\bibitem{Koch:1990:LRCC}
H. {Koch} and P. {J{\o}rgensen},  J. Chem. Phys.  \textbf{93}, 3333 (1990).

\bibitem{Stanton:1993:EOMCC}
J.F. {Stanton} and R.J. {Bartlett},  J. Chem. Phys.  \textbf{98}, 7029 (1993).

\bibitem{Shen:2009:BCCCexcitedstates}
J. {Shen} and S. {Li},  J. Chem. Phys.  \textbf{131}, 174101 (2009).

\bibitem{Seidler:2011:VCCResponseTheory}
P. {Seidler}, M. {Sparta} and O. {Christiansen},  J. Chem. Phys.  \textbf{134},
  054119 (2011).

\bibitem{Zimmerman:2017:iFCI}
P.M. {Zimmerman},  J. Chem. Phys.  \textbf{146}, 224104 (2017).

\end{thebibliography}
\end{document}